\documentclass[twocolumn,aps,prb,showpacs,preprintnumbers,amsmath,amssymb, superscriptaddress]{revtex4}

\usepackage{graphicx}
\usepackage{dcolumn}
\usepackage{bm}

\begin{document}


\title{Melting and Rippling Phenomena in Two Dimensional Crystals with localized bonding}


\author{D. J. Priour, Jr}
\affiliation{Department of Physics, University of Missouri, Kansas City, Missouri 64110, USA}
\author{James Losey}
\affiliation{Department of Physics, University of Missouri, Kansas City, Missouri 64110, USA}


\date{\today}

\begin{abstract}
We calculate Root Mean Square (RMS) deviations from equilibrium for atoms in a
two dimensional crystal with local (e.g. covalent) bonding between
close neighbors.  Large scale Monte Carlo calculations are in good
agreement with analytical results obtained in the harmonic approximation.
When motion is restricted to the plane, we find a slow (logarithmic)
increase in fluctuations of the atoms about their equilibrium positions
as the crystals are made larger and larger.
    We take into account fluctuations perpendicular to the lattice
plane, manifest as undulating ripples, by examining dual-layer systems
with coupling between the layers to impart local rigidly (i.e. as in sheets
of graphene made stiff by their finite thickness).  Surprisingly, we find a rapid
divergence with increasing system size in the vertical mean square deviations,
independent of the strength of the interplanar coupling.
   We consider an attractive coupling to a flat substrate, finding
that even a weak attraction significantly limits the amplitude
and average wavelength of the ripples.
    We verify our results are generic by examining a variety of
distinct geometries, obtaining the same phenomena in each case.
\end{abstract}

\maketitle


\section{Introduction}
Efforts to gain a quantitative microscopic understanding of melting have spanned more than a
century.  The Lindemann criterion developed in 1910~\cite{Lindemann}  describes melting in terms of the 
Root Mean Square (RMS) deviation from the atomic equilibrium positions.  Since long-range 
positional order stems from the periodic arrangement of atoms in crystalline solid, atomic
deviations that are comparable to the separation between atomic species could 
obscure the regularity of the underlying crystal lattice with a concomitant loss of 
positional order.  The Lindemann criterion specifies that melting has occurred if the 
RMS deviations reach on the order of a tenth of a lattice constant, and has proved 
to be a reasonably effective theory for three dimensional systems. 

The Lindemann analysis does not take into account correlations of the motions of 
neighboring atoms.  Correlations are more important at lower dimensions, and the process of 
melting is hence strongly dimensionally dependent.  While three dimensional crystals 
exhibit long-range order below certain temperatures, statistical fluctuations play a significant
role in one dimensional systems, precluding all but short-ranged local ordering for $T > 0$. 

The process of melting in two dimensions is more subtle, and is understood in the 
modern context to occur in more than one stage.  An initial continuous loss of positional order 
precedes the proliferation of lattice defects, which accumulate and eventually complete the melting process  
at sufficiently high temperatures by destroying even orientational order, where each atom has a fixed   
number of neighbors.   

   Thermally induced fluctuations in atomic positions
 can have an important effect on nano-engineered systems where structures 
may be on the atomic scale.  Atomic clusters or quantum ``dots'' are mesoscopic 
assemblies of atoms where the scale is confined in all directions.  Linear structures 
such as carbon nanotubes are essentially one dimensional objects (although having 
cross sections on the atomic scale) where the tube length may 
approach macroscopic scales.  Finally, two dimensional systems with nanoscale thickness 
such as covalently bonded graphene sheets are 
genuine monolayers with thicknesses on the atomic scale, but spanning macroscopic areas.

   The novel charge transport properties of graphene have been of intrinsic  
fundamental interest, and have also inspired scenarios for the use of graphene in 
semiconductor microprocessor applications. Technological uses for graphene will need a stable 
planar substrate for the implementation of nano-circuitry, and 
fundamental scientific research will also benefit from the minimization of the 
amplitude of random 
undulations in graphene layers.  

   We examine two dimensional crystals with properties that would generically be found in
 two dimensional covalently bonded crystals, including stiffness with respect to
displacements perpendicular to the plane of the sheet.  Although we do not consider temperature regimes  
capable of disrupting the lattice topology or number of neighbors for 
each atom (e.g. by thermal rupture of bonds 
between neighboring atomic species), we examine the loss of order caused by 
fluctuations of atomic positions about their equilibrium positions 
which nonetheless leave the bonding pattern intact. 

If the motion of particles comprising the crystal is confined to the plane of the 
lattice, the gradual loss of long-range crystalline order with increasing system size 
has been understood as being in some respects similar to the destruction of ferromagnetic ordering in 
the $X$-$Y$ model (the motion of the spins are confined to the plane with a ferromagnetic
coupling between them) by thermally excited 
spin waves.  Nevertheless, on a detailed level the two systems differ.  In the case of 
the $X$-$Y$ model, spin-spin correlation functions decay algebraically with the spatial separation 
between spins below the 
Kosterlitz-Thouless temperature for vortex unbinding.  On the other hand, 
the RMS deviation in atomic positions in two dimensional crystals has been described as logarithmically divergent (i.e. 
varying as $\log[ \alpha(T) L ]$ where $\alpha(T)$ is a temperature dependent 
parameter) for any finite temperature~\cite{Chaikin}.  

\begin{figure}
\includegraphics[width=.42\textwidth]{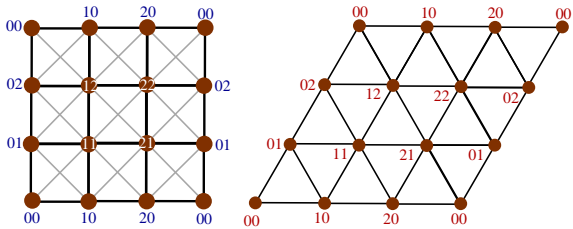}

\caption{\label{fig:Fig12} Square lattice extended coupling geometry with interactions with
nearest and next-nearest neighbors (left) and the periodic triangular lattice and labeling scheme (right)}
\end{figure}

In Section II, we discuss theoretical techniques and the system geometries under 
consideration.  Then, in section III   
we examine three dimensional lattices where we show directly for suitably rigid lattice 
geometries that 
the RMS deviations from equilibrium converge to a finite value in the 
thermodynamic limit, an anticipated property of three 
dimensional systems. Moreover, we determine a reference temperature threshold $T_{L}^{\textrm{3D}}$ where mean 
square fluctuations about equilibrium reach one tenth of a lattice constant, corresponding to the 
melting point according to the Lindemann criterion.  
This temperature
will serve as a point of reference in the examination of two dimensional systems where 
thermal fluctuations disrupt long-range order for any finite temperature.
However, although we find stable crystalline order in three dimensional geometries, we also 
discuss a significant caveat which applies for simple cubic lattices and other geometries which 
lack local stiffness.  To a great extent, the lattice geometries we report on are based on the 
two dimensional examples shown in Fig.~\ref{fig:Fig12}.  A square lattice pattern, and a triangular 
lattice structure are shown.  The former lacks inherent rigidity, but the square lattice gains
local rigidity through the activation of an extended coupling 
scheme in which both nearest neighbors and next-nearest neighbors interact.   
In the same way, a simple cubic lattice requires interactions between next-nearest
as well as nearest neighbors to resist thermal fluctuations and maintain long-range crystalline order.

By considering two geometries and appropriate three dimensional generalizations 
which differ in significant ways (i.e. one base on a square pattern and the other
assembled of triangles or tetrahedra joined at their corners), we identify generic  
thermodynamic characteristics common to both.

In Section IV, we examine two dimensional lattices such as those shown in Fig.~\ref{fig:Fig12} with 
motion confined to the crystal plane, finding the very slow (i.e. logarithmic) loss of crystalline 
order anticipated for two dimensional crystals.   
On the other hand, for a two dimensional crystal embedded in three dimensions, it is 
important to consider transverse perturbations tending to push atoms out of the plane.  We find 
that in the absence of binding to a substrate, two dimensional crystals are much less able to 
resist extraplanar distortions than fluctuations which are confined to the lattice plane.

In Section V, we examine dual-layer systems where coupling between the 
crystal planes imparts local stability with respect to extra-planar variations of 
atomic positions in a caricature of physical systems (e.g. sheets of graphene) which
have a finite thickness, and would be imbued with local stiffness.  
We find for two distinct locally rigid dual-layer geometries
similar rapid divergences of mean square displacements as the  
crystal is made larger, corresponding to 
thermally induced rippling of the crystal, and scaling linearly with the size of 
the system.  Analysis of the density of vibrational states 
reveals that the length scale of the random undulations increases with the size of 
the system with strong long wavelength contributions.
On the other hand coupling to a flat substrate, however weak, places an 
asymptotic upper bound on the ripple amplitudes and also limits the average 
wavelength of thermally induced undulations.

\section{Calculation methods and Monte-Carlo Simulation Results}

We examine thermodynamic properties (e.g. the mean square deviations of 
atoms about equilibrium positions) for crystals with short range bonding
in the regime where bonds 
remain intact and thermally induced lengthening and shortening of bonds is small 
relative to the unperturbed, or equilibrium, bond length. With individual
bonds varying only slightly in length, it is appropriate to model the bonds as
harmonic potentials so the couplings between neighboring atoms are effectively 
treated as springs connecting the two particles.  It is important to note that 
although we neglect anharmonic effects from the bonds, the noncollinearity of 
bonds in the crystal geometry may in principle introduce anharmonic terms in the Hamiltonian.
Nevertheless, at temperatures near and below the melting point, many scenarios 
are amenable to the harmonic approximation where the neglect of anharmonicities (whether 
intrinsic or geometric) has a small impact on the accuracy of the calculation.  Analytical results obtained in the 
context of the harmonic approximation are validated in the cases we consider 
by good agreement with Monte-Carlo 
calculations where the anharmonic characteristics of the bonding stemming from peculiarities 
of the lattice geometry are rigorously 
taken into account.  

The Hamiltonian is given by 
\begin{equation}
\mathcal{H} = \frac{1}{2} \sum_{i=1}^{N} \sum_{j = 1}^{m_{i}} \frac{K_{ij}}{2} \left( l_{ij} - l_{ij}^{0} \right)^{2}
\label{eq:eq1}
\end{equation}
where $l_{ij}^{0}$ is the equilibrium energetically favored bond length and $l_{ij}$ is the 
instantaneous separation between atoms $i$ and $j$.  The outer sum is over the 
atoms in the (finite) crystal, and the inner sum is over the neighbors associated with the atom indexed 
by the label $i$. The additional factor of 1/2 is included to compensate for double counting of bonds.  
The constant $K_{ij}$ is the second derivative of the interatomic potential $V_{ij}(r)$ at the equilibrium
 separation $l_{ij}^{0}$.  

We develop the harmonic approximation directly from the bond length  
$l_{ij} = \sqrt{(x_{i} - x_{j})^{2} + (y_{i} - y_{j})^{2} + (z_{i} - z_{j})^{2} }$
For the $x$ coordinates, it is convenient to write, for example, $x_{i} = x_{i}^{0} + \delta_{i}^{x}$ where
$x_{i}^{0}$ is the equilibrium coordinate and $\delta_{i}^{x}$ is the shift about equilibrium.
We operate in the same way for the $y$ and $z$ coordinates, finding 
\begin{align}
l_{ij} = \sqrt{ \begin{array}{c} (\Delta_{ij}^{0x} + \delta_{i}^{x} - \delta_{j}^{x})^{2}
+ (\Delta_{ij}^{0y} + \delta_{i}^{y} - \delta_{j}^{y})^{2} \\ + 
(\Delta_{ij}^{0z} + \delta_{i}^{z} - \delta_{j}^{z})^{2} \end{array}},
\end{align}
where $\Delta_{ij}^{0x} \equiv (x_{i}^{0} - x_{j}^{0})$, 
$\Delta_{ij}^{0y} \equiv (y_{i}^{0} - y_{j}^{0})$, and 
$\Delta_{ij}^{0z} \equiv (z_{i}^{0} - z_{j}^{0})$.
One may develop the harmonic approximation by expanding terms such as $(l_{ij} - l_{ij}^{0})^2$ to 
quadratic order in the shift differences $(\delta_{i}^{x} - \delta_{j}^{x})$, 
$(\delta_{i}^{y} - \delta_{j}^{y})$, and $(\delta_{i}^{z} - \delta_{j}^{z})$. 
The result will be
$(l_{ij} - l_{ij}^{0}) \approx \left[ \hat{\Delta}_{ij} \cdot \left ( \vec{\delta}_{i} - \vec{\delta}_{j} \right ) \right]^{2}$,
where $\hat{\Delta}_{ij}$ is a unit vector formed from $\vec{\Delta}_{ij} = (\Delta_{ij}^{0x}, 
\Delta_{ij}^{0y}, \Delta_{ij}^{0z})$.The terms $\vec{\delta}_{i}$ and $\vec{\delta}_{j}$ are vector 
atomic displacements such that, e.g., 
$\vec{\delta}_{i} = (\delta_{i}^{x},\delta_{i}^{y},\delta_{i}^{z})$. 
A salient characteristic of the bond energy is its dependence on the differences of 
the coordinate shifts (e.g.  $\delta_{i}^{x} - \delta_{j}^{x}$ for the $x$ direction) instead of  
$\delta_{i}^{x}$, $\delta_{i}^{y}$, and $\delta_{i}^{z}$ by themselves, a condition which under 
many circumstances permits the neglect of anharmonicities due to bond non-collinearity. 

In the harmonic approximation, the lattice energy due to deviations from equilibrium positions
will be
\begin{align}
\mathcal{H}^{\mathrm{Har}} = \frac{1}{2} \sum_{i=1}^{N} \sum_{j = 1}^{m_{i}} \frac{K_{ij}}{2} 
\left[ \hat{\Delta}_{ij} \cdot \left ( \vec{\delta}_{i} - \vec{\delta}_{j} \right ) \right]^{2}
\end{align}
On expanding, one obtains a quadratic expression mixing the displacements
\begin{align}
\mathcal{H}^{\mathrm{Har}} \! = \! \sum_{i=1}^{N} \sum_{j=1}^{m_{i}}
\! \tfrac{K_{ij}}{4} \! \left[ \begin{array}{ccc} \delta_{i}^{x} & 
\delta_{i}^{y} & \delta_{i}^{z} \end{array} \! \right] \! \! \!
\left[ \! \begin{array}{ccc} a_{xx} & a_{xy} & a_{xz} \\ 
a_{yx} & a_{yy} & a_{yz} \\ a_{zx} & a_{zy} & a_{zz} \end{array} \! \right] \! \! \! \! 
\left[ \! \begin{array}{c} \delta_{j}^{x} \\
\delta_{j}^{y} \\ \delta_{j}^{z} \end{array} \! \right] 
\end{align}
Diagonalizing the appropriate matrix yields 3$N$ eigenvectors, taken to be normalized. 
Each of the set of 3N eigenvectors has a component for the individual degrees of 
freedom in the crystal lattice, permitting the lattice Hamiltonian to be 
written in decoupled form as 
 \begin{align}
 \mathcal{H}^{\mathrm{Har}} = \frac{K}{2} \sum_{\alpha = 1}^{3N} \lambda_{\alpha}  c_{\alpha}^{2} 
 \end{align} 
with eigenvector expansion coefficients $c_{\alpha}$ and eigenvalues
$\lambda_{\alpha}$; the parameter $K$ is the ``primary'' harmonic constant, which is taken to be the nearest neighbor 
intra-planar coupling constant in schemes, such as extended models with multiple coupling constants.  The eigen-modes are   
independently excited by thermal fluctuations, and thermodynamic equilibrium 
observables may be calculated by 
evaluating Gaussian integrals.  As an example, the thermally averaged mean square fluctuation per 
atomic species $\langle \delta_{\textrm{RMS}} \rangle$ is (first moments of the coordinate shifts such as 
$\langle \delta_{i}^{x} \rangle$ vanish in the thermal average and do not appear in the expression below)
\begin{align}
\langle \delta_{\textrm{RMS}} \rangle^{2} = \tfrac{1}{N} \sum_{i=1}^{N} \langle (\delta_{i}^{x})^{2} 
+ (\delta_{i}^{y})^{2} + (\delta_{i}^{z})^{2} \rangle  
\end{align}
 Indexing the eigenvectors with the label $\alpha$ and noting, e.g., 
 that $\delta_{i}^{x} = \displaystyle{\sum_{\alpha = 1}^{3N}} c_{\alpha} v_{\alpha}^{ix}$, we see that 
 the total square of the instantaneous fluctuations per particle is
 \begin{align}
 \delta_{\mathrm{RMS}} = \tfrac{1}{N}  \sum_{i=1}^{N} \!  
  \sum_{\alpha=1}^{3N} \sum_{\alpha^{'}=1}^{3N} \left [ c_{\alpha} c_{\alpha^{'}} ( v_{\alpha}^{ix} 
 v_{\alpha^{'}}^{ix}  +  v_{\alpha}^{iy} v_{\alpha^{'}}^{iy}    +  
v_{\alpha}^{iz} v_{\alpha^{'}}^{iz}) \right ]
 \end{align}
 In calculating the thermal average the term $c_{\alpha} c_{\alpha^{'}}$ 
 will be as often negative as positive when $\alpha \neq \alpha^{'}$, and there will only be a non-zero contribution 
 to $\langle \delta_{\mathrm{RMS}} \rangle^{2} $ if   
 $\alpha = \alpha^{'}$. Hence, the double sum enclosed in square brackets will collapse to a single sum, and the  
 calculation is reduced to the thermal average 
 \begin{align} 
 \langle \delta_{\mathrm{RMS}}  \rangle^{2} = \tfrac{1}{N}  \sum_{\alpha=1}^{3N} 
 \langle  c_{\alpha}^{2} \rangle \sum_{i=1}^{N}\left[ (v_{\alpha}^{ix})^{2} + (v_{\alpha}^{iy})^{2} + 
 (v_{\alpha}^{iz})^{2} \right]  
 \end{align}
The eigenvector normalization condition gives
\begin{align}
 \sum_{i=1}^{N} \left[ (v_{\alpha}^{ix})^{2} + 
(v_{\alpha}^{iy})^{2} + (v_{\alpha}^{iz})^{2} \right] = 1,
\end{align}
and hence $\langle \delta_{\mathrm{RMS}}  \rangle^{2}$ appears simply as 
\begin{align}
\langle \delta_{\mathrm{RMS}}  \rangle^{2} = \tfrac{1}{N}  \sum_{\alpha=1}^{3N} 
\langle ( c_{\alpha})^{2} \rangle
\end{align}
The partition function $Z$ may be calculated with the aid of $\int_{-\infty}^{\infty} e^{-\sigma q^{2}} dq  = 
(\pi /\sigma)^{1/2}$, and one has a product of decoupled Gaussian integrals, which may be written as
\begin{align}
Z = \prod_{\alpha=1}^{3N} \int_{-\infty}^{\infty} e^{-K \beta \lambda_{\alpha} c_{\alpha}^{2}/2 } d c_{\alpha}  
\label{Eq:eq100}
\end{align}
with $\beta = 1/k_{\mathrm{B}}$, $k_{\mathrm{B}}$ the Boltzmann constant, and the temperature $T$ is 
given in Kelvins.
For the sake of convenience, units are chosen such that the lattice constant $a$ is equal to unity, and a reduced
temperature is defined with $t \equiv k_{\mathrm{B}} T/K$.  Evaluating the integrals in the product given in 
Eq.~\ref{Eq:eq100} yields for $Z$
\begin{align}
Z = \prod_{\alpha=1}^{3N} \left ( \frac{2 \pi t}{\lambda_{\alpha}} \right )^{1/2}
\end{align} 
The thermally averages mean square displacement may be written in terms of a thermal logarithmic derivative of $Z$, 
and in particular, one finds
\begin{align}
\langle \delta_{\mathrm{RMS}} \rangle^{2} = t^{2} \frac{d}{dt} \textrm{Ln}(Z) = \sum_{\alpha = 1}^{3N} \lambda_{\alpha}^{-1} t
\end{align}
Hence, the thermally averaged mean square deviation from equilibrium may be written as the square root of a sum over eigenvalue 
reciprocals.
\begin{align}
\langle \delta_{\mathrm{RMS}} \rangle = t^{1/2} \sqrt{ \sum_{\alpha = 1}^{3N} \lambda_{\alpha}^{-1}}
\end{align}
Zero eigenvalues would lead to a diverging expression, but eigenvalues which are strictly equal to zero 
are artifacts of periodic boundary conditions, correspond to 
global translations of the crystal lattice, and are excluded from the sum.
The dependence on reduced temperature consists of a $t^{1/2}$ factor.  To concentrate on characteristics  
specific to a lattice geometry and its coupling scheme, as well as trends with respect to system size $L$, the normalized mean 
square displacement $\delta_{\mathrm{RMS}}^{{n}}$ will often be discussed in lieu of the full   
temperature dependent quantity. 

In the case of a periodic
regular crystal lattice, it is useful
exploit translational invariance, which will lead to exact expressions for the
vibrational mode eigenstates and frequencies for periodic crystals (or at the very least
yielding a small matrix which may be diagonalized analytically or by numerical means if necessary)
if atomic displacements are written in terms of the corresponding Fourier components.

Using Monte Carlo calculations to sample thermodynamic quantities incorporates anharmonic effects  
in a rigorous manner, providing a means of assessing the validity of the harmonic approximation.  
We employ the Metropolis technique~\cite{Metropolis} to introduce random displacements
and sample the distribution corresponding to thermal equilibrium.  
We follow the standard Metropolis prescription, where an attempted random 
displacement with an associated energy shift $\Delta E$ is 
accepted with probability $e^{-\Delta E/k_{\mathrm{B}} T}$ if $\Delta E > 0$ and the Monte Carlo move is
invariably accepted for cases in which $\Delta E < 0$.

\begin{figure}
\includegraphics[width=.48\textwidth]{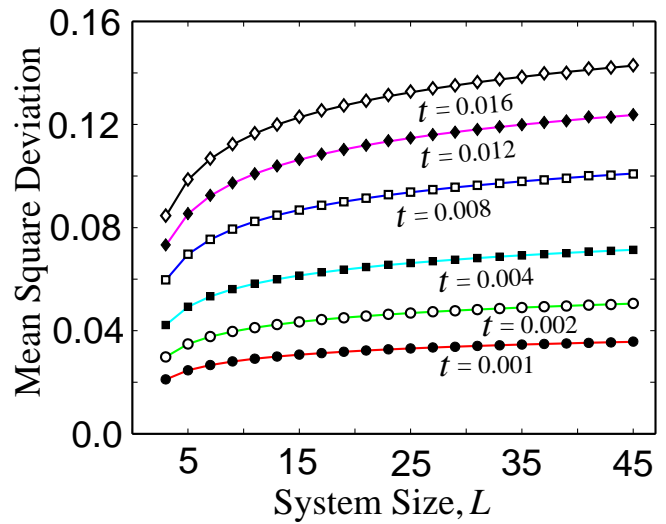}
\caption{\label{fig:Fig1} Graph of the RMS fluctuation about
equilibrium versus systems size $L$ for various values of the reduced temperature $t$ for the
square lattice with next-nearest neighbor couplings.  The
solid lines are analytic results obtained in the harmonic approximation, and 
symbols are results from Monte Carlo calculations.}
\end{figure}

\begin{figure}
\includegraphics[width=.48\textwidth]{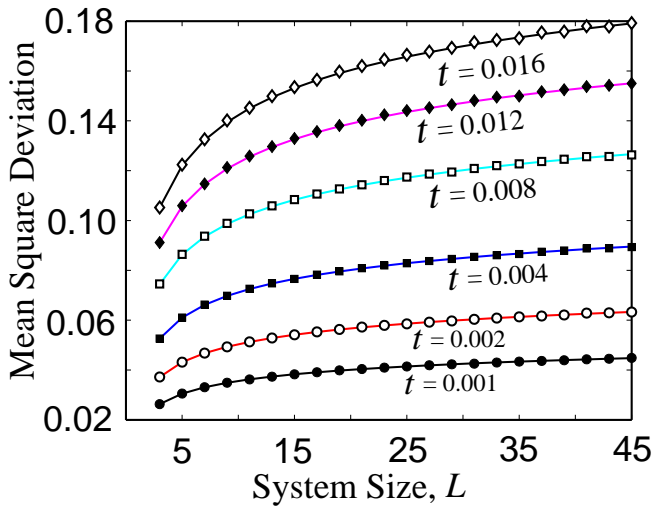}
\caption{\label{fig:Fig2} Graph of the RMS fluctuation about 
equilibrium versus systems size $L$ for various values of the reduced temperature $t$ for the 
triangular lattice.  The 
solid lines are analytic results obtained in the harmonic approximation, and 
symbols are results from Monte Carlo calculations.} 
\end{figure}

In calculating thermodynamic quantities, we operate in terms of Monte Carlo sweeps where a sweep,
on average, consists of an attempt to move each atom in the crystal with the acceptance of the move
subject to the Metropolis condition.   In the calculations, the sampling of thermodynamic quantities  
is postponed until the completion of the first 25\% of the total number of sweeps to eliminate bias from the initial 
conditions, which are not typical thermal equilibrium configurations for the system.  To reduce 
errors due to statistical fluctuations in the Monte Carlo simulation and obtain several digits of 
accuracy in the results, we conduct at least $5 \times 10^{5}$ sweeps.  
Figure~\ref{fig:Fig1} (for the square lattice with an extended coupling scheme) and Figure~\ref{fig:Fig2} 
(for the triangular lattice geometry) show mean square deviation curves for 
various temperatures ranging from temperatures an order of magnitude smaller than 
$T_{\mathrm{3D}}^{\mathrm{L}}$ to temperatures on par with the Lindemann criterion result for the 
melting temperature of the corresponding three dimensional system.  The solid lines correspond to 
analytical results, while the symbols are RMS values obtained with Monte Carlo calculations.  

The curves show very good agreement between the Monte Carlo data and analytical results over a wide 
range of temperatures and system sizes, 
and deviations are primarily mild statistical errors  
(on the order of one part in $10^{3}$) in the Monte Carlo calculations.

\begin{figure}
\includegraphics[width=.35\textwidth]{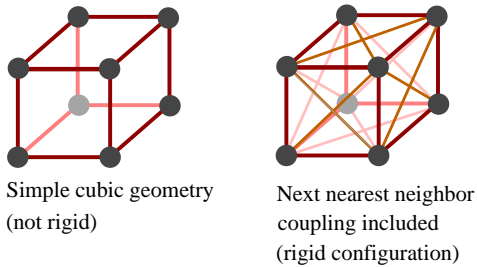}

\caption{\label{fig:Fig3} Illustration of the simple cubic, nonrigid
structure and rigidity gained by incorporating next nearest-neighbor couplings
as shown in the image to the right.}  
\end{figure}

\section{Rigid and Non-Rigid Three Dimensional lattices} 

To establish a temperature scale for the two dimensional systems, where long-range crystalline 
order is not expected to exist at temperatures above 0K, we first examine three dimensional lattices, which may exhibit 
long-range positional order at finite temperature if the lattice is suitably rigid.
As a preliminary step, we perform an analysis similar to the Lindemann treatment where an atom in a simple 
cubic geometry is coupled to six nearest neighbors.
Since we do not take into account the motion of 
neighboring atoms, we take their displacements to be zero; 
certainly the excursions of neighboring atoms would average to zero, although to be 
precise, one would need to take into account cooperative effects of the atomic motions of 
the neighbors.

The lattice energy has the form 
\begin{align}
\hat{h}_{\mathrm{L}}^{\mathrm{3D}} = \frac{K}{2} \left ( \begin{array}{c} \Delta_{x_{+}}^{2} + \Delta_{x_{-}}^{2} + \Delta_{y_{+}}^{2} + 
\Delta_{y_{-}}^{2} + \Delta_{z_{+}}^{2} + \Delta_{z_{-}}^{2}  \end{array} \right) 
\end{align}
where, for example, $\Delta_{x_{+}}$ is the shift in length of the bond to the nearest neighbor in 
the positive $\hat{x}$ direction.
Applying the harmonic approximation and taking the atomic 
shifts to be $\left \{ \delta_{x}, \delta_{y}, \delta_{z} \right \}$, the energy becomes
\begin{align}
\hat{h}_{\mathrm{L}}^{\mathrm{3D}} = K \left[ \delta_{x}^{2} + 
\delta_{y}^{2} + \delta_{z}^{2}  \right]
\end{align} 
In the calculation of $\delta_{\mathrm{RMS}}$, the partition function has the form
\begin{align}
Z = \int^{\infty}_{-\infty} \int^{\infty}_{-\infty} \int^{\infty}_{-\infty} e^{-\tfrac{\delta_{x}^{2} + 
\delta_{y}^{2} + \delta_{z}^{2}}{t}} d \delta_{x} d \delta_{y} d \delta_{z}
\end{align}
From the Gaussian integration, we find $Z =   
(\pi t)^{3/2}$.  The RMS displacement will be $\langle r^{2} \rangle^{1/2}$ where 
\begin{align}
\langle r^{2} \rangle = t^{2} \tfrac{d}{dt} \textrm{Ln} (Z)  = 3t/2 
\end{align}
Hence, the thermally averaged mean square shift is $(3t/2)^{1/2}$.  The Lindemann criterion places 
the melting temperature at a temperature high enough that the mean square deviation $\delta_{\mathrm{RMS}}$
reaches a tenth of a lattice constant, 
which corresponds to $t_{\mathrm{L}}^{\mathrm{3D}} 
= \tfrac{2}{3}(10^{-2})$, a reduced temperature on the order of $0.01$.
 
If correlations among atoms are taken into account, next-nearest neighbor couplings become 
crucial to imparting local stiffness and maintaining long-range crystalline order. 
To see how rigidity is an important factor, we calculate the RMS displacements for a simple cubic 
lattice where only couplings between nearest neighbors are taken into account.  The energy stored in 
the lattice will be 
\begin{align}
E = \frac{K}{2} \sum_{i,j,k = 0}^{n-1} \left[ \! \! \begin{array}{c} (\delta_{i+1jk}^{x} - \delta_{ijk}^{x})^{2} + 
(\delta_{ij+1k}^{y} - \delta_{ijk}^{y})^{2} \\ + (\delta_{ijk+1}^{z}-\delta_{ijk}^{z})^{2} \end{array} \! \! \right]
\end{align}
where a periodic geometry is assumed, and the counting factor of $1/2$ does not appear since the 
sum has been constructed to avoid redundancies.  If we use the transformations 
\begin{align}
\delta_{ijk}^{x} = \sum_{k_{x}k_{y}k_{z}} \delta_{\bf{k}}^{x} e^{I(k_{x}i + k_{y}j + k_{z}k)}
\end{align}
where $I$ is the imaginary unit, and similar expressions are used for $\delta_{ijk}^{y}$ and 
$\delta_{ijk}^{z}$.
In terms of the Fourier components, the lattice energy may be written as
\begin{align} 
E = \frac{K}{2} \! \! \sum_{k_{x},k_{y},k_{z}} \left [ \! \begin{array}{c} (1- \cos k_{x}) | \delta_{\bf{k}}^{x}|^{2} 
\\ + ( 1  - \cos k_{y}) | \delta_{\bf{k}}^{y}|^{2} \\ + (1 - \cos k_{z}) | \delta_{\bf{k}}^{z} |^{2}
\end{array} \! \right  ] 
\end{align}
The $x$, $y$, and $z$ degrees of freedom $\delta_{\bf{k}}^{x}$, $\delta_{\bf{k}}^{y}$, and 
$\delta_{\bf{k}}^{z}$ automatically decouple.
The normalized 
mean square deviation is $\sqrt{\sum_{\alpha} \lambda_{\alpha}^{-1}}$, where the sum is restricted to non-zero eigenvalues.
We identify three eigenvalues, $\lambda_{\bf{k}}^{(1)}  = 2(1 -\cos k_{x})$, 
$\lambda_{\bf{k}}^{(2)} = 2 (1 - \cos k_{y})$, and $\lambda_{\bf{k}}^{(3)} = 2 (1 - \cos k_{z})$
for each wave vector $\left \{ k_{x},k_{y},k_{z} \right \}$
As can be seen in Figure, the mean square fluctuation about equilibrium positions grows very rapidly with increasing system size.
The divergence in the RMS displacements is a consequence of the lack of rigidity of the simple cubic geometry, 
which facilitates the destruction of long range crystalline order by thermal fluctuations. 
However, next-nearest neighbor couplings make the lattice rigid, and are very effective in 
suppressing fluctuations about equilibrium and establishing long-range crystalline order 
for the simple cubic lattice.  

\begin{figure}
\includegraphics[width=.49\textwidth]{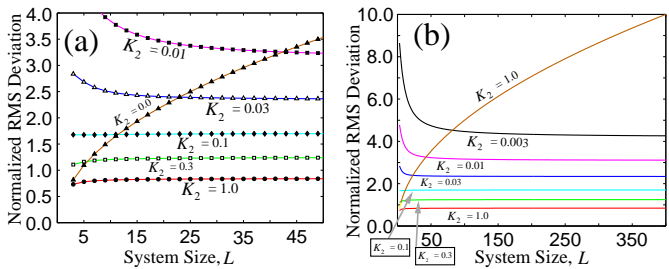}

\caption{\label{fig:Fig4} RMS displacements graphed with system size $L$ for various values of 
$K_{2}$, expressed in units of the nearest-neighbor coupling $K_{1}$. Panel (a) shows a closer view 
of the $\delta_{\mathrm{RMS}}$ curves over a smaller range of system sizes, and panel (b) is a graph with a  
broader range of system sizes included in the plot.}
\end{figure}

\begin{figure}
\includegraphics[width=.49\textwidth]{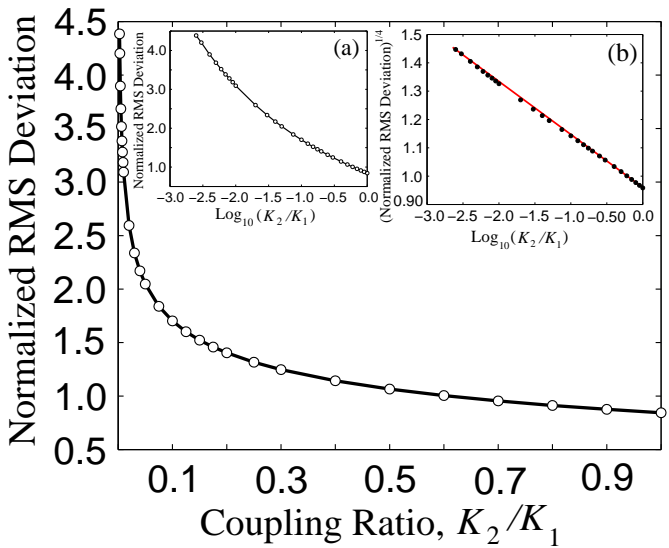}

\caption{\label{fig:Fig5} RMS displacements graphed as a function of the ratio of next nearest 
neighbor to nearest neighbor coupling, $K_{2}/K_{1}$.  Inset (a) is a graph of the normalized RMS
deviation with respect to $\log_{10} (K_{2}/K_{1})$, and inset (b) shows the RMS fluctuations 
raised to the 1/4 power versus $\log_{10} (K_{2}/K_{1})$.}
\end{figure}

\begin{figure}
\includegraphics[width=.35\textwidth]{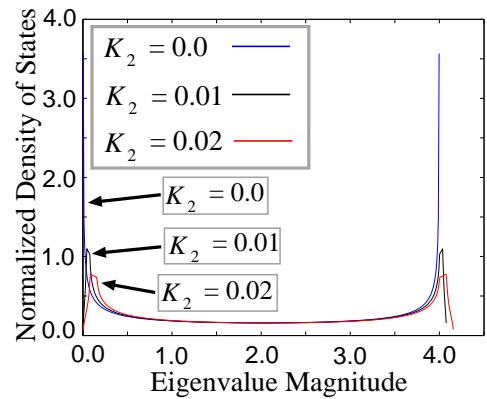}

\caption{\label{fig:Fig6} Normalized Eigenvalue Density of States for the simple cubic system
for an extended coupling scheme with $K_{2} = \left \{ 0.0, 0.01, 0.02 \right \}$ 
with a sampling of $2.4 \times 10^{8}$
eigenvalues.}
\end{figure}

The structure of the eigenvalue density states 
profile has informative characteristics particular to the lattice geometry from which it is obtained, and the 
density of states is calculated for many of the systems we report on.  We 
achieve the thermodynamic in a genuine sense by not restricting $k_{x}$, 
$k_{y}$, and $k_{z}$ to discrete values as is done for finite systems. 
The density of states is built up by Monte Carlo sampling in which 
the wave-vector components are each generated independently from a uniform random 
distribution.
To obtain good statistics,  
at least on the order of $2 \times 10^{8}$ eigenvalues are sampled in constructing the 
DOS.  The same Monte Carlo sampling procedure is used to calculate the $\delta_{\mathrm{RMS}}$ 
values shown in Fig.~\ref{fig:Fig5}, and thereby completely remove any bias from finite size effects.  The density of states corresponding
to the simple cubic system (shown in the graph in Fig.~\ref{fig:Fig6}) is consistent with the divergence of the RMS 
fluctuations with increasing system size.
The bimodal structure is sharply peaked in the low and high eigenvalue regimes, with the 
former contributing to the steady rise of $\delta_{\mathrm{RMS}}$ with increasing system size $L$.

\begin{figure}
\includegraphics[width=.45\textwidth]{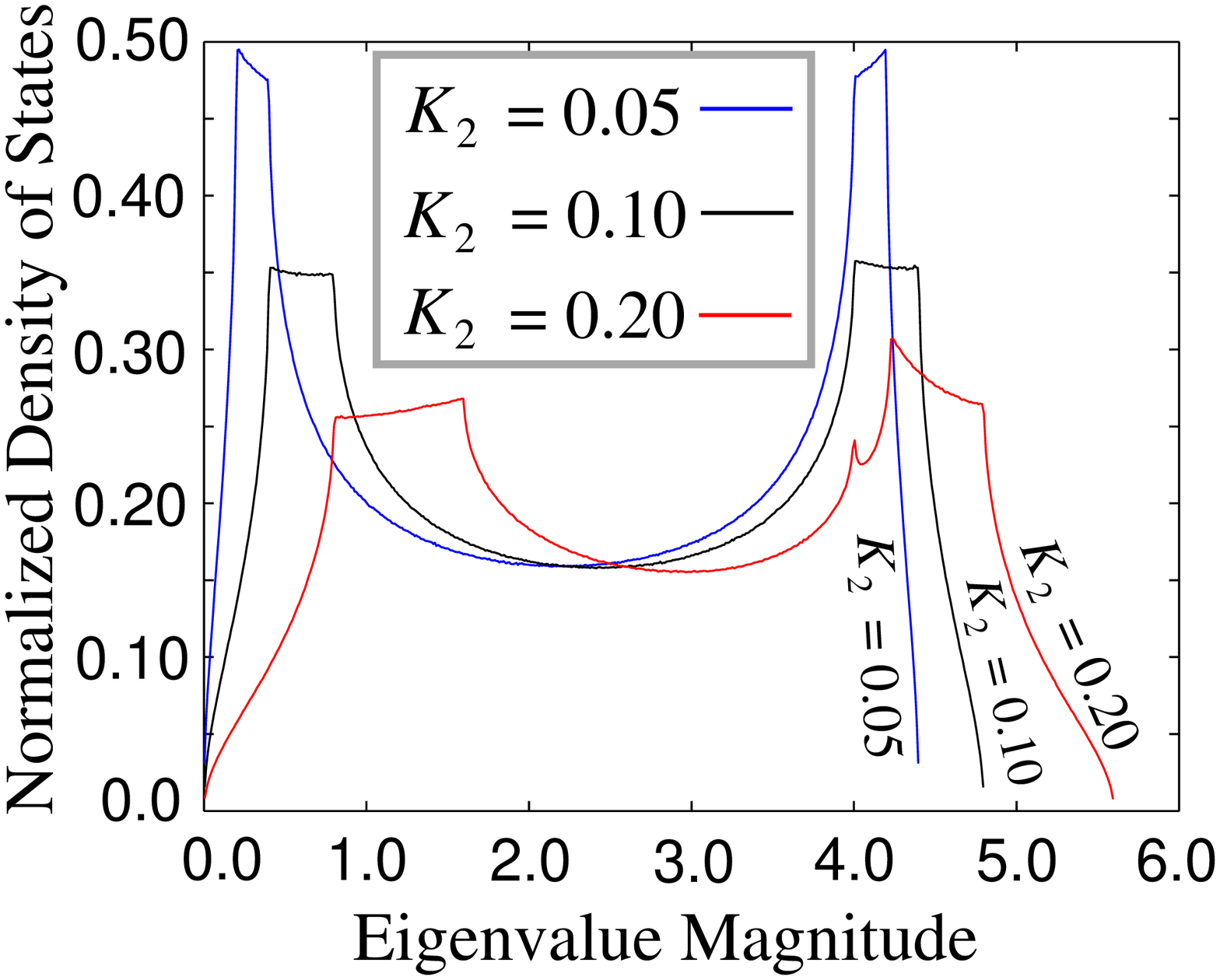}

\caption{\label{fig:Fig7} Normalized Eigenvalue Density of States for the simple cubic system
for an extended coupling scheme with $K_{2} = \left \{0.05, 0.10, 0.20 \right \}$ with a 
sampling of $2.4 \times 10^{8}$
eigenvalues.}
\end{figure}

\begin{figure}
\includegraphics[width=.45\textwidth]{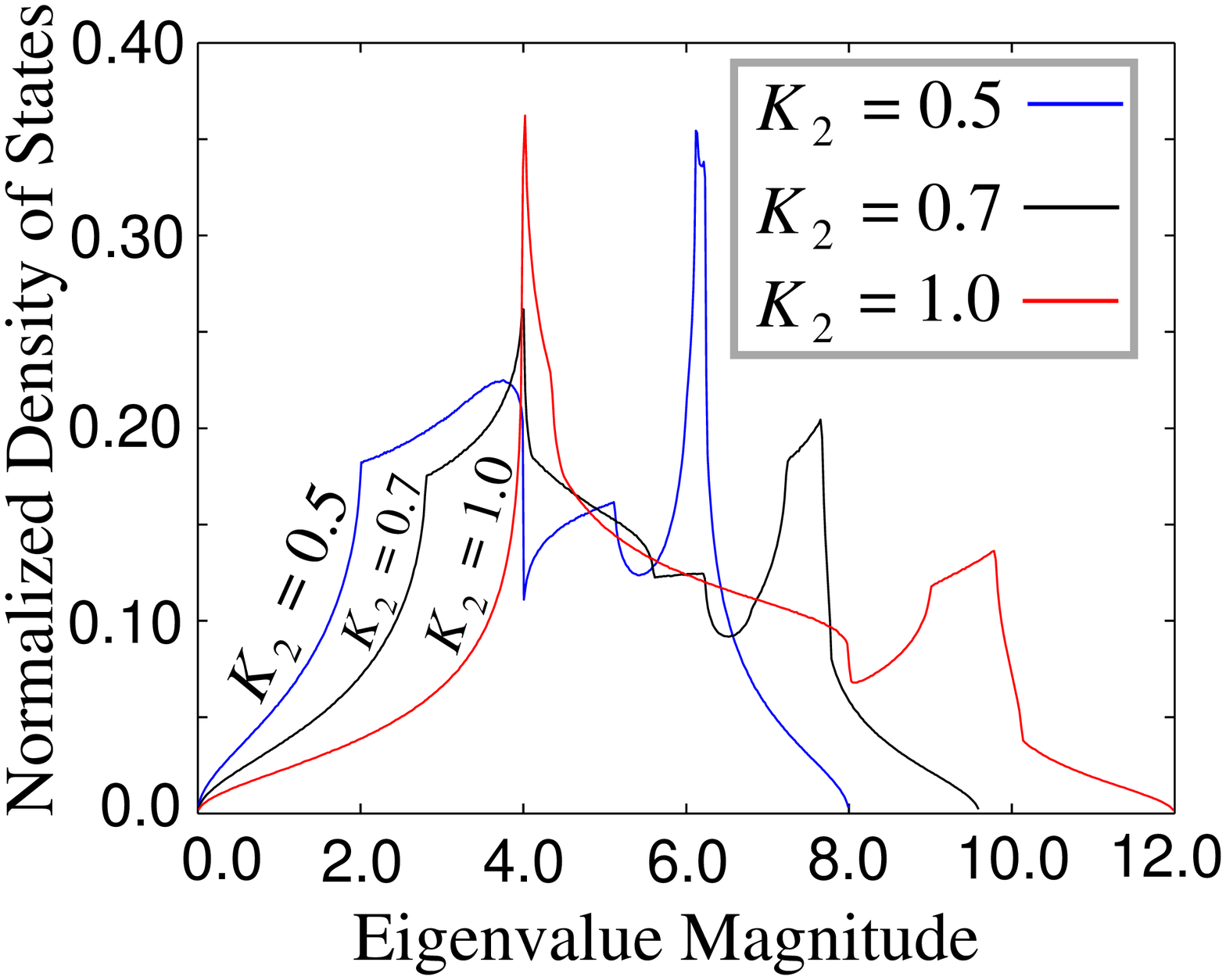}

\caption{\label{fig:Fig8} Normalized Eigenvalue Density of States for the simple cubic system
with coupling for an extended coupling scheme with $K_{2} = 
\left \{ 0.5, 0.7, 1.0 \right \}$ with a sampling of $2.4 \times 10^{8}$
 eigenvalues.}
\end{figure}

In the extended coupling scheme in the simple cubic geometry, the energy stored in the lattice is
\begin{align}
 E  \! = \!
\frac{K_{1}}{2} \! \! \! \! \sum_{i,j,k=0}^{n-1}  \! \left( \! \! \! \begin{array}{c}
\left \{ \! \! \!  \begin{array}{l} [ \hat{x} \! \cdot \! (\vec{\delta}_{i+1jk} \! - \! \vec{\delta}_{ijk})]^{2} 
\! + \! [\hat{y} \! \cdot \! (\vec{\delta}_{ij+1k} \! - \! \vec{\delta}_{ijk})]^{2} \\  \! + \!
[ \hat{z} \! \cdot \! (\vec{\delta}_{ijk+1} \! - \! \vec{\delta}_{ijk}) ]^{2} \end{array} \! \! \right \}   \\
\! \! \! +  \kappa_{2} \! \left \{  \! \! \! \begin{array}{l} \left[ \frac{1}{\sqrt{2}}(\hat{x} + \hat{y}) \cdot 
(\vec{\delta}_{i+1j+1k} - \vec{\delta}_{ijk}) \right]^{2} \! + \! \\
\left[ \frac{1}{\sqrt{2}}( \hat{x} \!- \!\hat{y}) \! \cdot \! (\vec{\delta}_{i+1j-1k} \! - \! \vec{\delta}_{ijk} ) \right]^{2} \! \! +  \! \\
\left[ \frac{1}{\sqrt{2}}( \hat{y} \!+ \!\hat{z}) \! \cdot \! (\vec{\delta}_{ij+1k+1} \!- \! \vec{\delta}_{ijk} ) \right]^{2} \! \! + \! \\
\left[ \frac{1}{\sqrt{2}}( \hat{y} \! - \! \hat{z}) \! \cdot \! (\vec{\delta}_{ij+1k-1} \!- \! \vec{\delta}_{ijk} ) \right]^{2} \! \! + \! \\
\left[ \frac{1}{\sqrt{2}}( \hat{x} \! + \! \hat{z}) \! \cdot \! (\vec{\delta}_{i+1jk+1} \!- \! \vec{\delta}_{ijk} ) \right]^{2} \! \! + \! \\
\left[ \frac{1}{\sqrt{2}}( \hat{x} \! - \! \hat{z}) \! \cdot \! (\vec{\delta}_{i+1jk-1} \!- \! \vec{\delta}_{ijk} ) \right]^{2}
\end{array}
\! \! \! \!  \right \} 
\end{array}
\! \! \! \! \right)
\end{align} 
where $K_{1}$ is the coupling to nearest neighbors, $K_{2}$ is the coupling to next-nearest neighbors, and
$\kappa_{2} \equiv K_{2}/K_{1}$ is the ratio of the next-nearest and nearest neighbor coupling constants.
In terms of Fourier components, one has 
\begin{align}
E \! = \! \frac{K_{1}}{2} \! \!  \sum_{\bf{k}}  \! \!
\left( \! \! \!
\begin{array}{l}
 \left \{ \! \! \begin{array}{l} \! (2 \! - \! 2 \cos k_{x}) \lvert \delta_{\bf{k}}^{x} \rvert^{2} \! + \! 
(2 - 2 \cos k_{y}) \lvert \delta_{\bf{k}}^{y} \rvert ^{2} \\ \! + \! (2 \! - \! 2 \cos k_{z}) \lvert \delta_{\bf{k}}^{z} \rvert ^{2} \end{array} 
\! \! \right \} \! + \!  \\ 
\\
\! \kappa_{2} \! \left \{ \! \! \begin{array}{l} \left [ 2 \! -  \!  \cos k_{x} \cos k_{y} \!- \! 
\cos k_{x} \cos k_{z} \right ] \! \lvert \delta_{\bf{k}}^{x} \rvert^{2} \! + \!\\
\left[ 2 \! - \! \cos k_{x} \cos k_{y} \! - \! \cos k_{y} \cos k_{z} \right] 
\! \lvert \delta_{\bf{k}}^{y}\rvert^{2} \! + \! \\
\left[ 2 \! - \! \cos k_{y} \cos k_{z} \! - \! \cos k_{x} \cos k_{z} \right]
\! \lvert \delta_{\bf{k}}^{z} \rvert^{2} \! + \! \\
\sin k_{x} \sin k_{y} ( \delta_{\bf{k}}^{x} \delta_{\bf{k}}^{*y} 
+ \delta_{\bf{k}}^{*x} \delta_{\bf{k}}^{y}) + \\
\sin k_{y} \sin k_{z} ( \delta_{\bf{k}}^{y} \delta_{\bf{k}}^{*z}
+ \delta_{\bf{k}}^{*y} \delta_{\bf{k}}^{z}) + \\
\sin k_{z} \sin k_{x} ( \delta_{\bf{k}}^{z} \delta_{\bf{k}}^{*x}
+ \delta_{\bf{k}}^{*z} \delta_{\bf{k}}^{x})
\end{array} \! \! \! 
\right \}
\end{array} \! \! \! \! \!
\right )
\end{align}
with $\bf{k}$ indicating the wave-vector with components $k_{x}$, $k_{y}$, and $k_{z}$, and 
again $\kappa_{2} = K_{1}/K_{2}$.
The eigenvalues are hence obtained by diagonalizing the 3$\times$3 matrix
\begin{align}
2 \! \! \left[ \! \! \! 
\begin{array}{ccc}
 \left( \! \! \begin{array}{c} 2 \! - \! \cos k_{x} \cos k_{y}\\ \! - \! \cos k_{x} \cos k_{z} \end{array}  \! \! \! \right)  \! \! & \! \!
 \kappa_{2} \sin k_{x} \sin k_{y}   \! \! \! & \! \! \!  \kappa_{2} \sin k_{x} \sin k_{z}  \\  \\
 \kappa_{2} \sin k_{x} \sin k_{y}  \! \! \! & \! \! \! \left( \! \! \begin{array}{c} 2 \! - \! \cos k_{y} \cos k_{z} \\ 
\! - \! \cos k_{z} \cos k_{x} \end{array} \! \! \! \right ) \! \! \! & \! \! \! \kappa_{2} \sin k_{y} \sin k_{z}  \\ \\
\kappa_{2} \sin k_{z} \sin k_{x} \! \! \! & \! \! \! \kappa_{2} \sin k_{z} \sin k_{y} \! \! \! & \! \! \!  
\left( \! \! \begin{array}{c} 2 \! - \! \cos k_{z} \cos k_{y} \\ \! - \! 
 \cos k_{z} \cos k_{x} \end{array} \! \! \! \right) 
\end{array} \!  \! \!
\right ]
\end{align}
Although solving the cubic characteristic equation will yield analytical expressions for the eigenvalues, 
the result is cumbersome, and we instead use standard algorithms for the diagonalization of a symmetric
matrix to efficiently obtain the eigenvalues numerically.

The eigenvalues determined in this manner are used to calculate the means square atomic fluctuations, and the 
results are shown in Figure~\ref{fig:Fig4}, where $\delta_{\mathrm{RMS}}$ is graphed with respected to $L$ for a range of 
the next to nearest neighbor coupling strength ratio $\kappa_{2} = K_{2}/K_{1}$.  Whereas the mean square displacement 
steadily rises with system size when $\kappa_{2} = 0$ (i.e with only nearest-neighbor couplings active), 
the curves behave very differently for nonzero $\kappa_{2}$, ultimately saturating with increasing $L$.
The stabilization of $\delta_{\mathrm{RMS}}$ in the thermodynamic limit indicates the presence of intact 
long-range crystalline order.  In the case of $\kappa_{2} = 0$, the mean square deviation steadily 
diverges with increasing $L$.  The same divergence with only nearest-neighbor interactions taken into a
account occurs whether one is considering the simple cubic structure, a square lattice, or a 
linear chain.  Hence, in some lattice geometries, having a three dimensional structure may be insufficient to stabilize long-range 
order if an extended coupling scheme is not taken into consideration.

By switching on and varying the strength of the next-nearest coupling $K_{2}$,
one sees the appearance of long-range crystalline order as the cubic system is
made increasingly rigid.  In Fig.~\ref{fig:Fig5} the mean square displacement is shown graphed
versus the coupling ratio $K_{2}/K_{1}$.  The tendency for atoms to be driven from
their positions in the lattice does increase as $K_{2}$ is shut off, but the divergence
occurs at a slow rate.  Inset (a) is a graph of $\delta_{\textrm{RMS}}$ versus the
logarithm of $K_{2}/K_{1}$.  While the concavity of the curve indicates a somewhat more
rapid than logarithmic divergence, a semi-logarithmic plot of $\delta_{\mathrm{RMS}}$
(i.e. as shown in inset (b) of Fig.~\ref{fig:Fig5}) shows an approximately linear scaling of
$\delta_{\mathrm{RMS}}^{1/4}$ with the logarithm of the system size, still a relatively
slow divergence, albeit somewhat more rapid than a simple logarithmic divergence.
Hence, the next-nearest neighbor couplings in the extended coupling simple cubic model are
very effective in restoring long-range crystalline order.

Trends in the eigenvalue density of states profile with increasing $K_{2}/K_{1}$
 to next-nearest neighbors are shown in
in Fig.~\ref{fig:Fig6}, Fig.~\ref{fig:Fig7}, Fig.~\ref{fig:Fig8}.  The almost immediate retreat of the low and high frequency peaks
toward the center is consistent with the effectiveness of an extended coupling scheme in stabilizing 
long-range crystalline order even for very small values of the ratio $\kappa_{2} = K_{2}/K_{1}$.  
The DOS profile has a simple structure for small $\kappa_{2} < 0.1$, while 
intermediate $\kappa_{2}$ values are associated with a richer density of 
states curve which changes rapidly as the coupling ratio is increased further.

\begin{figure}
\includegraphics[width=.5\textwidth]{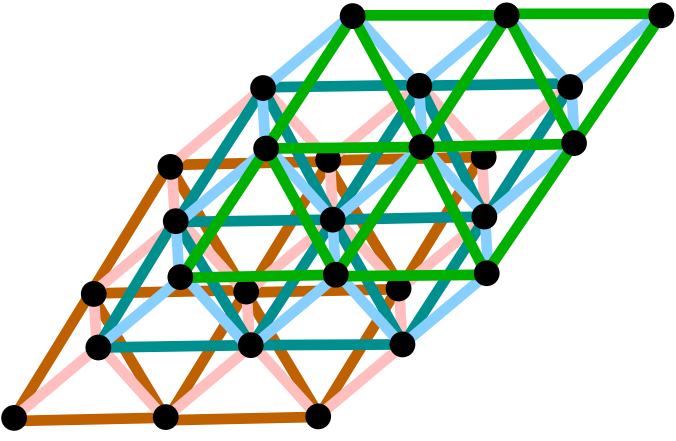}

\caption{\label{fig:Fig11} Illustration of the tetrahedral lattice geometry}
\end{figure}

As in the case of the cubic lattice with the extended coupling scheme, we may 
calculate the lattice energy in the harmonic approximation for the tetrahedral lattice, which is
\begin{align}
E = \! \frac{K}{2} \! \! \sum_{i,j,k=0}^{n-1} \! \! \left (    \! \!
\begin{array}{l}
\left[ \vec{x} \! \cdot \! (\vec{\delta}_{i+1jk} \! - \! \vec{\delta}_{ijk}) \right]^{2} + \\
\left[ \! \left (\! \frac{1}{2} \hat{x} \! + \! \frac{\sqrt{3}}{2} \hat{y} \! \right ) \! 
\cdot \! (\vec{\delta}_{ij+1k} \! - \! \vec{\delta}_{ijk}) \! \right]^{2} \! + \! \\
\left[ \! \left ( \! \frac{1}{2} \hat{x} \! - \! \frac{\sqrt{3}}{2} \hat{y} \! \right ) 
\! \cdot \! (\vec{\delta}_{i+1j-1k} \! - \! \vec{\delta}_{ijk}) \! \right]^{2} \! + \! \\ 
\left[ \! \left ( \! \frac{1}{2} \hat{x} \! + \! \frac{1}{2\sqrt{3}} \hat{y} \! + \! \sqrt{\frac{2}{3}} \hat{z} \! \right ) \! \cdot \!
(\vec{\delta}_{ijk+1} \! - \! \vec{\delta}_{ijk} ) \! \right]^{2} \! + \! \\
\left[ \! \left ( \! -\frac{1}{2} \hat{x} \! + \! \frac{1}{2 \sqrt{3}} \hat{y} \! + \! \sqrt{\frac{2}{3}} \hat{z} \! \right ) \! \cdot \! 
(\vec{\delta}_{i-1jk+1} \! - \! \vec{\delta}_{ijk}) \! \right]^{2} \\
\! + \! \left[ \left ( -\frac{1}{\sqrt{3}} \hat{y} \! + \! \sqrt{\frac{2}{3}} \hat{z} \! \right ) 
\cdot (\vec{\delta}_{ij-1k+1} \! - \! \vec{\delta}_{ijk}) \! \right]^{2}
\end{array}  \! \! \! \right )
\end{align}
where there is only one coupling constant $K$ since bonds are considered between nearest neighbors only,
the tedrahedral geometry being intrinsically rigid,
and we have used the fact that the altitude of a tetrahedron is $\sqrt{\tfrac{2}{3}}$ times the lattice 
constant.  The energy may be expressed in terms of Fourier components, and one has the task of  
diagonalizing the $3 \times 3$ matrix
\begin{align}
\left[ \! \! \! \! \!
\begin{array}{ccc}
 \left( \! \! \! \! \begin{array}{c} 4 - 2 \cos k_{x}  - \\ \frac{1}{2} (\cos k_{y} \!  + \! \cos k_{z}) \\
- \frac{1}{2} \cos (k_{y} \! - \! k_{x}) \\ - \frac{1}{2} \cos (k_{z}\!  - \! k_{y} ) 
\end{array} \! \! \! \! \right)  
\! \! \! \! \! \! \! &
 \tfrac{\sqrt{3}}{2} \! \! 
\left( \! \! \! \begin{array}{c} \cos (k_{y} \! - \! k_{x} ) \\ - \cos k_{y} \end{array} \! \! \! \right) \! \!
\! \! \!  &  
\sqrt{\tfrac{2}{3}} \! \! \left ( \! \! \! \begin{array}{c} \cos (k_{z} - k_{x}) \\ -     
\cos k_{z} \end{array} \! \! \! \right)  \\  \\
 \tfrac{\sqrt{3}}{2} \! \! 
\left ( \! \! \! \begin{array}{c} \cos (k_{y} \! - \! k_{x} ) \\ - \cos k_{y} \end{array} \! \! \! 
\right )  \! \! \! \! \! \! \! & \left( \! \! \! \! \begin{array}{c} 4 - \tfrac{3}{2} \cos k_{y} \\
- \tfrac{3}{2} \cos (k_{y} \! - \! k_{x} ) \\ - \tfrac{1}{6} \cos k_{z} - \\
 \tfrac{2}{3} \cos (k_{z} \! - \! k_{y} ) \\ - \tfrac{1}{6} \cos (k_{z}\!  - \! k_{z} ) 
  \end{array} \! \! \! \! \right) \! \! \! \! \! 
& \tfrac{1}{\sqrt{3}} \! \! \left( \! \! \! \begin{array}{c} 2 \cos (k_{z} \!  - \! k_{y} ) \\
- \cos k_{z} - \\ \cos ( k_{z} \! - \! k_{x} ) \end{array} \! \! \! \right)    \\ \\
\sqrt{\tfrac{2}{3}} \! \!  \left ( \! \! \! \begin{array}{c} \cos (k_{z} \! - \! k_{x}) \\ -
\cos k_{z} \end{array} \! \! \! \right) \! \! \! \! \! \! \! &  
\tfrac{1}{\sqrt{3}} \! \! \left( \! \! \!  \begin{array}{c} 2 \cos (k_{z} \! - \! k_{y} ) \\
- \cos k_{z} - \\ \cos ( k_{z} \! - \! k_{x} ) \end{array} \! \! \! \right) \! \! \! \! \! &  
\tfrac{4}{3} \! \! \left( \! \! \! \! \begin{array}{c} 
3 - \cos k_{z} - \\ \cos (k_{z} \!  - \! k_{x})  - \\ 
 \cos (k_{z} \! - \! k_{y} )   \end{array} \! \! \! \! \right )
\end{array} \! \! \! \!
\right ]
\end{align}

\begin{figure}
\includegraphics[width=.49\textwidth]{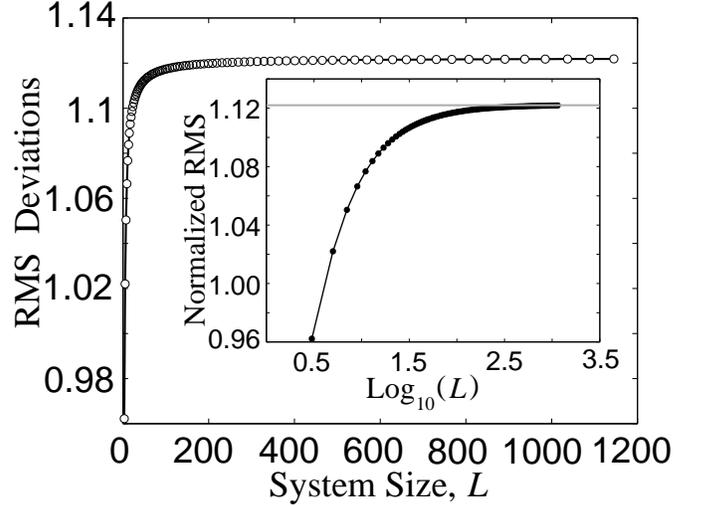}

\caption{\label{fig:Fig9} Normalized root mean square (RMS) deviation shown versus $\log_{10}L$ for the
three dimensional tetrahedral crystal.  The inset
is a graph of the normalized RMS deviations, again plotted with respect to $\log_{10}L$,
with the horizontal line indicating the extrapolated $\delta_{\mathrm{RMS}}$ in the thermodynamic 
limit.}
\end{figure}

The three dimensional tetrahedral lattice is locally stiff even with only nearest-neighbor couplings taken 
into account, and the rigidity inherent in the tetrahedral lattice geometry is sufficient to preserve 
long-range crystalline order, as may be seen in Figure~\ref{fig:Fig9} which displays the normalized mean square 
deviation versus the system size $L$.  The inset is a semi-logarithmic plot with the horizontal axis extending
over three decades of system sizes.  The saturation of the normalized RMS displacement with increasing $L$ 
is evident in both of the graphs, and in the thermodynamic limit, is in the vicinity of $1.12$.  With 
temperature dependence included, one will have $\delta_{\mathrm{RMS}} = 1.12 t^{1/2}$.  Hence, the Lindemann 
criterion would give $t_{3\mathrm{D}}^{\mathrm{L}} = 0.0080$, compatible with the previous estimate which    
neglected correlations of the atomic displacements from equilibrium.

\begin{figure}
\includegraphics[width=.49\textwidth]{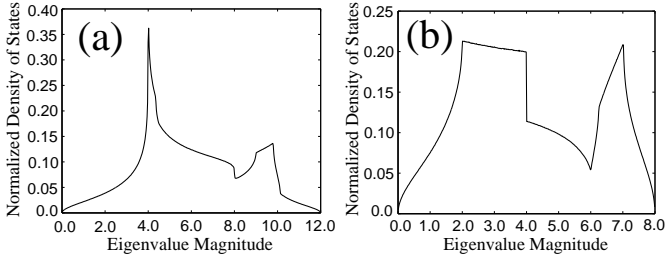}
\caption{\label{fig:Fig10} Normalized Eigenvalue Density of States for the                              
cubic crystal with nearest and next-nearest neighbor couplings (a),
and the density of states profile for the tetrahedral lattice (b).}          
\end{figure}

In inset (a) of Fig.~\ref{fig:Fig10}, the density of states is shown for the simple cubic lattice geometry
with the extended coupling scheme, and for the tetrahedral lattice in inset (b).  For both systems,
while other details of the density of states profiles differ,
the curves tend swiftly to zero in the small eigenvalue regime, a hallmark of intact long
range crystalline order in rigid three dimensional lattices.

\section{Intraplanar Motion}

We first examine the case where motion perpendicular to the plan is 
forbidden, and atomic deviations from equilibrium are confined 
to the lattice plane.  We consider various geometries, but first 
we consider a square (effectively a face-centered system), illustrated in Figure~\ref{fig:Fig1}, where
coupling to the four next nearest neighbors is taken into  
account.  We then consider a locally rigid triangular lattice where each atom interacts
with six nearest neighbors.
In both the face-centered square and triangular systems, we find a logarithmic divergence with increasing system
size in the mean square fluctuations about equilibrium.

For the periodic square geometry with the coupling scheme extended to next-nearest neighbors, 
the lattice energy to quadratic order is  
\begin{align}
E = \frac{K}{2} \! \! 
\sum_{i,j=0}^{n-1}  \left ( \! \! \begin{array}{l} \left [\hat{x} \! \cdot \! \left( \vec{\delta}_{i+1j}^{x} \!-\!  
\vec{\delta}_{ij}^{x} \right) \! \right]^{2} \! + \! \left[ \hat{y} \! \cdot \! 
\left( \vec{\delta}_{ij+1}^{y} \! - \! \vec{\delta}_{ij}^{y} \right) \! \right]^{2}  
\\ \! + \! \left [ \! \tfrac{1}{\sqrt{2}} ( \hat{x} \! + \! \hat{y} ) \! \cdot \! 
 \left( \vec{\delta}_{i+1j+1}^{x} \! - \! \vec{\delta}_{ij}^{x} \right) \! \right]^{2} \\
\! + \!  \left[ \! \tfrac{1}{\sqrt{2}} ( \hat{x} \! - \! \hat{y} ) \! \cdot \! 
  \left( \vec{\delta}_{i+1j-1}^{x} \! - \! \vec{\delta}_{ij}^{x} \right) \! \right ]^{2} 
 \end{array} \! \! \right )
\end{align}
Operating in reciprocal space, one diagonalizes the $2 \times 2$ matrix
\begin{align}
\left[  \! \! \begin{array}{ll} \left( \! \! \begin{array}{c} 2 -  \cos k_{x} \\- \cos k_{x} \cos k_{y}  
 \end{array} \! \! \right) &  \sin k_{x} \sin k_{y} \\ 
\sin k_{x} \sin k_{y} & 
\left( \! \! \begin{array}{c} 2 -  \cos k_{y} \\- \cos k_{x} \cos k_{y}
 \end{array} \! \! \right )  \end{array} \! \! \right ]
\end{align}
yielding the eigenvalues
\begin{align}
\lambda^{\pm}_{\bf{k}} = \left( \begin{array}{c} 4 - \cos k_{x} - \cos k_{y} - 2 \cos k_{x} \cos k_{y}  \end{array} 
\right) \\ \nonumber
\pm \sqrt{\begin{array}{c} (\cos k_{x} - \cos k_{y} )^{2} +  4 \sin^{2} k_{x} \sin^{2} k_{y} 
 \end{array} }
\end{align}

In the case of the triangular lattice with six fold coordination, one may also obtain
analytical expressions for the mean square deviations.  In real space, the
harmonic approximation for the energy stored in the lattice is
\begin{align}
E = \frac{K}{2} \sum_{i,j=0}^{n-1} \left ( \! \! \begin{array}{l} \left[ \hat{x} \! \cdot \! 
\left( \vec{\delta}_{i+1j} - \vec{\delta}_{ij} \right) \right]^{2} + \\
\left [(\frac{1}{2}\hat{x} + \frac{\sqrt{3}}{2} \hat{y} ) \! \cdot \!  \left( \vec{\delta}_{ij+1} - \vec{\delta}_{ij} \right) \right]^{2} + \\
\left [(\frac{1}{2}\hat{x} - \frac{\sqrt{3}}{2} \hat{y} ) \! \cdot \! \left( \vec{\delta}_{i+1j-1} - \vec{\delta}_{ij} \right) \right]^{2}
\end{array} \! \! \right ) 
\end{align} 
Expressing the displacements in terms of Fourier components, one decouples the $x$ and $y$ degrees of 
freedom by diagonalizing the matrix
\begin{align}
\left [ \! \! \! \begin{array}{cc} \left( \! \! \begin{array}{c} 3 \! - \! 2 \cos k_{x} \! - \! \\ 
\tfrac{1}{2}\cos k_{y} \! - \! \tfrac{1}{2} \cos [k_{y} \! - \! k_{x}] \end{array} \! \! \right ) \! \! & 
\tfrac{\sqrt{3}}{2} \left ( \cos [k_{y} \! - \! k_{x}] \! - \! \cos k_{y} \right) \\ \\ 
\tfrac{\sqrt{3}}{2} \left ( \cos [k_{y} \! - \! k_{x}] \! - \! \cos k_{y} \right) \! \! & \left( \! \! 
\begin{array}{c} 3 \! - \! \tfrac{3}{2} \cos k_{y} \\ \! - \! \tfrac{3}{2} \cos [k_{y} \! - \! k_{x}] \end{array} \! \! \right) 
\end{array} \! \! \! \right]
\end{align}
yields the eigenvalues 
\begin{align}
\! \! \! \! \! \lambda_{\bf{k}}^{\pm} = \left[ 3  - \cos k_{x} - \cos k_{y} - 
\cos (k_{y} - k_{x} ) \right] \\ \nonumber 
\pm \sqrt{\begin{array}{c} \cos^{2} k_{x} \! + \! \cos^{2} k_{y} \! + \! \cos^{2} (k_{y} \!  - \! k_{x})
\!   - \!  
\cos k_{x} \cos k_{y}  \\ \! - \!  \cos k_{y} \cos (k_{y} \! - \! k_{x}) \! - \! 
\cos (k_{y} \! - \! k_{x} ) \cos k_{x} \end{array}}
\end{align}
For convenience in comparison with the analytical 
results in the harmonic approximation, we consider periodic boundary conditions in the crystal plane.  
We have also examined anchored lattices, where atoms at the periphery are prevented from moving,  
while those in the interior are free to move.
For both the free and fixed boundary conditions, as in the three dimensional case, 
we obtain qualitatively similar results, and 
the same physical phenomena.

\begin{figure}
\includegraphics[width=.49\textwidth]{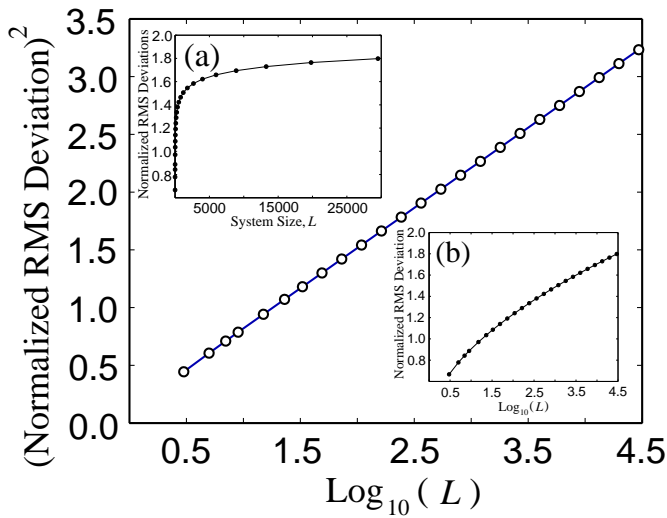}

\caption{\label{fig:Fig13}
Square of the normalized root mean square (RMS) deviation shown versus $\log_{10}L$ for the 
square lattice system with extended couplings.  The solid line
encompassing the open circular symbols is a strictly linear fit.  Inset (b)    
is a semi-logarithmic graph of the normalized RMS deviations, plotted with respect to $\log_{10}L$.}
\end{figure}

\begin{figure}
\includegraphics[width=.49\textwidth]{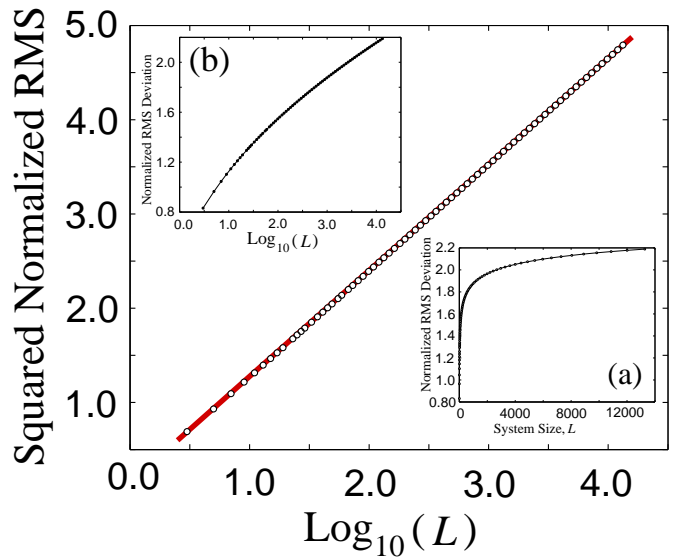}

\caption{\label{fig:Fig14} Square of the normalized root mean square (RMS) deviation shown versus $\log_{10}L$ for 
the triangular lattice.  The solid line 
encompassing the open circular symbols is a strictly linear fit.  Inset (a) is a standard plot of
the RMS deviation with respect to system size $L$, while inset (b) 
is a semi-logarithmic graph of the normalized mean square deviations, plotted with respect to $\log_{10}L$.}
\end{figure}

In Fig.~\ref{fig:Fig13} and Fig.~\ref{fig:Fig14}, the normalized mean square deviation $\delta^{{n}}_{\mathrm{RMS}}$
is graphed with respect to the system size $L$  for the square lattice in the extended scheme and the triangular lattice, 
respectively.  The overall behavior of the mean square deviations from equilibrium is qualitatively the same for both 
lattice geometries.  In both cases, the main graph is semi-logarithmic with $(\delta_{\mathrm{RMS}}^{{n}})^{2}$
on the ordinate.  The traces are linear to a very good approximation for all regimes of $L$ (i.e. for 
small, moderate, and large) shown, and the linearity is maintained for four decades 
of system sizes ranging from several to on the order of a few times $10^{4}$ lattice 
constants.  

In Fig.~\ref{fig:Fig13} and Fig.~\ref{fig:Fig14}, inset (a) is a standard plot, and the apparent saturation of the 
$\delta_{\mathrm{RMS}}^{n}$ curve is a hallmark of the slow loss of long-range crystalline order best seen on a 
semi-logarithmic graph.  Inset (b) in Fig.~\ref{fig:Fig13} and Fig.~\ref{fig:Fig4} contains as semi-logarithmic 
plot with $\delta_{\mathrm{RMS}}^{{n}}$ [ instead of $(\delta_{\mathrm{RMS}}^{{n}})^{2}$ ] on the ordinate axis.
The curves plotted in this manner are not linear, and it is evident that the divergence of the fluctuations about 
equilibrium is actually somewhat slower than logarithmic; instead, it is $(\delta_{\mathrm{RMS}}^{{n}})^{2}$ which  
scales as $\textrm{Ln}(L)$.

\begin{figure}
\includegraphics[width=.49\textwidth]{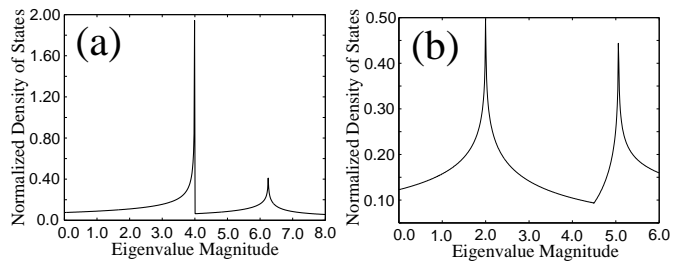}

\caption{\label{fig:Fig15} Normalized Eigenvalue Density of States for the face 
centered square lattice
with motion confined to the lattice plane, depicted in panel (a), and for the 
triangular lattice in panel (b).} 
\end{figure}

As in the case of the three dimensional systems, it is informative to examine the density of states,
shown in the graph of Fig.~\ref{fig:Fig15} for the square lattice in the extended coupling scheme in panel (a) and
the triangular lattice in panel (b) of Fig.~\ref{fig:Fig15}.  Again, while details of the density of states profiles shown are
peculiar to the lattice under consideration, the behavior in the regime of low eigenvalues is quite similar, and
both curves tend to a finite value instead of dropping swiftly to zero as in the density of states for the
rigid three dimensional lattices.  The failure of the density of states to vanish in the small  
eigenvalue limit contributes to the slow divergence of $\delta_{\mathrm{RMS}}$ in $L$.

\section{Extra-planar Motion}

The locally stiff face-centered square and triangular lattices show the 
anticipated slow logarithmic divergence in
system size.  However since laboratory    
systems often are not vertically constrained, it 
is important to examine a scenario where motion perpendicular to the plane of the lattice 
may be considered.  There is an important difficulty with single layer systems, in 
that motion perpendicular to the plane is not hindered since there are no restraining bonds with a 
directional component transverse to the plane of the layer. 

However, by considering dual-layer geometries, it is possible to incorporate local
stiffness with respect to perturbations that would push atoms above or below the lattice.
We examine analogs of the simple cubic lattice, where we again use an extended coupling scheme
to create local stiffness.  On the other hand, we also consider a dual-layer tetrahedral lattice. 
Although the two lattice geometries achieve local stiffness in different ways, the similarities we find in thermodynamic 
behavior of the mean square atomic fluctuations suggest these characteristics would appear in the generic case as well. 

\begin{figure}
\includegraphics[width=.5\textwidth]{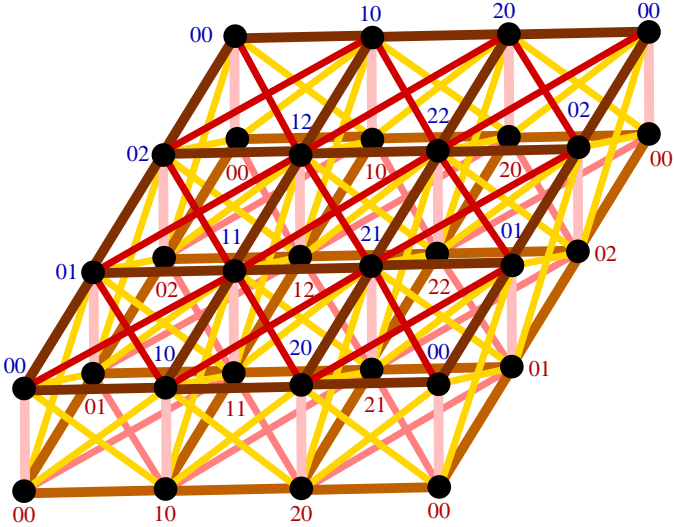}

\caption{\label{fig:Fig16} Illustration of the periodic dual-layer square lattice 
with nearest and next-nearest neighbor coupling and labeling scheme; blue indices refer to 
the upper layer, while red indices pertain to the lower plane.}
\end{figure}

Fig.~\ref{fig:Fig16}  illustrates the structure of the dual-layer square lattice with an 
extended coupling scheme; the additional couplings between next nearest neighbors impart local 
stiffness to the system with respect to perturbations perpendicular to the planes of the square 
lattices.

\begin{figure}
\includegraphics[width=.12\textwidth]{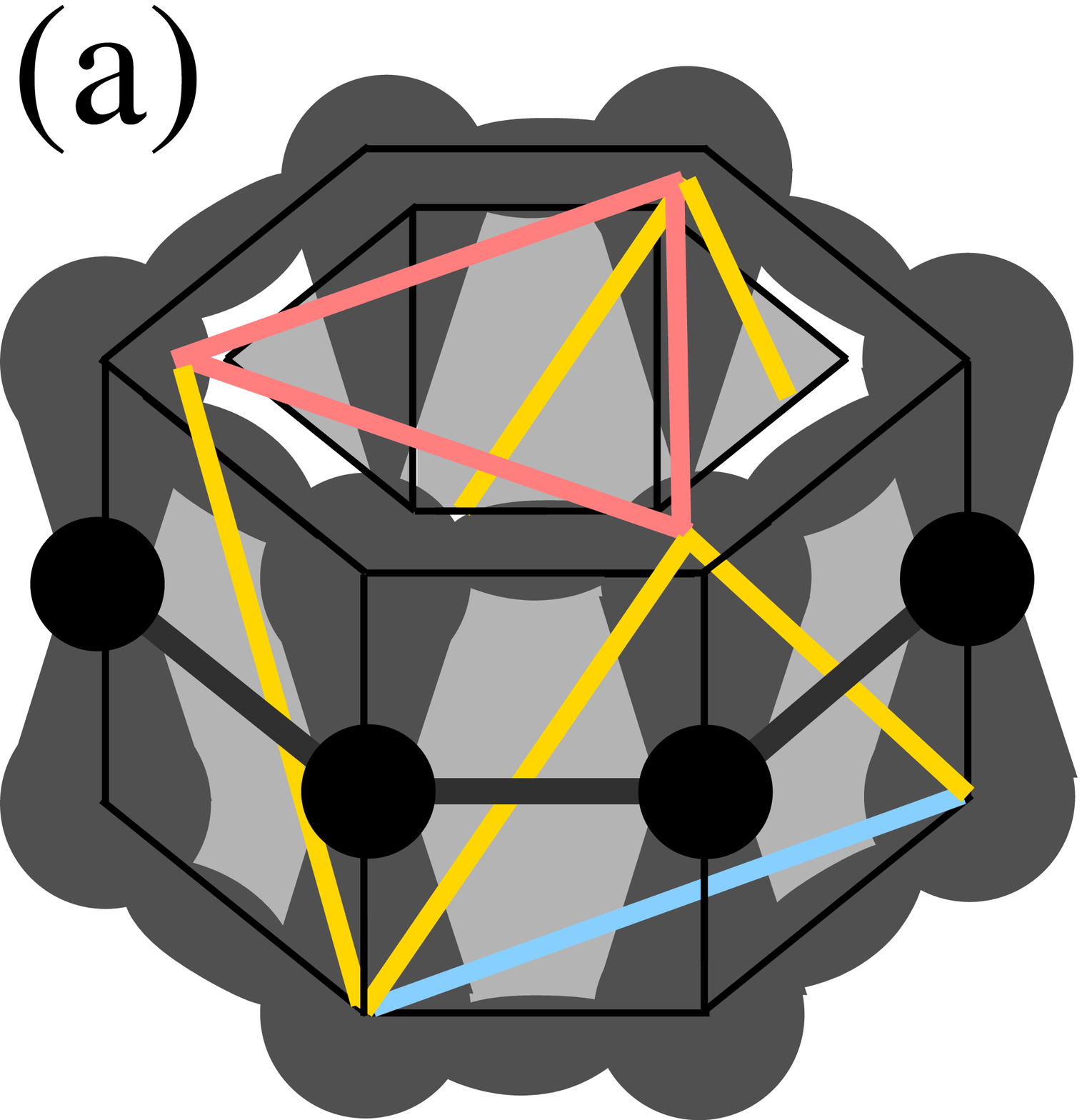}
\includegraphics[width=.30\textwidth]{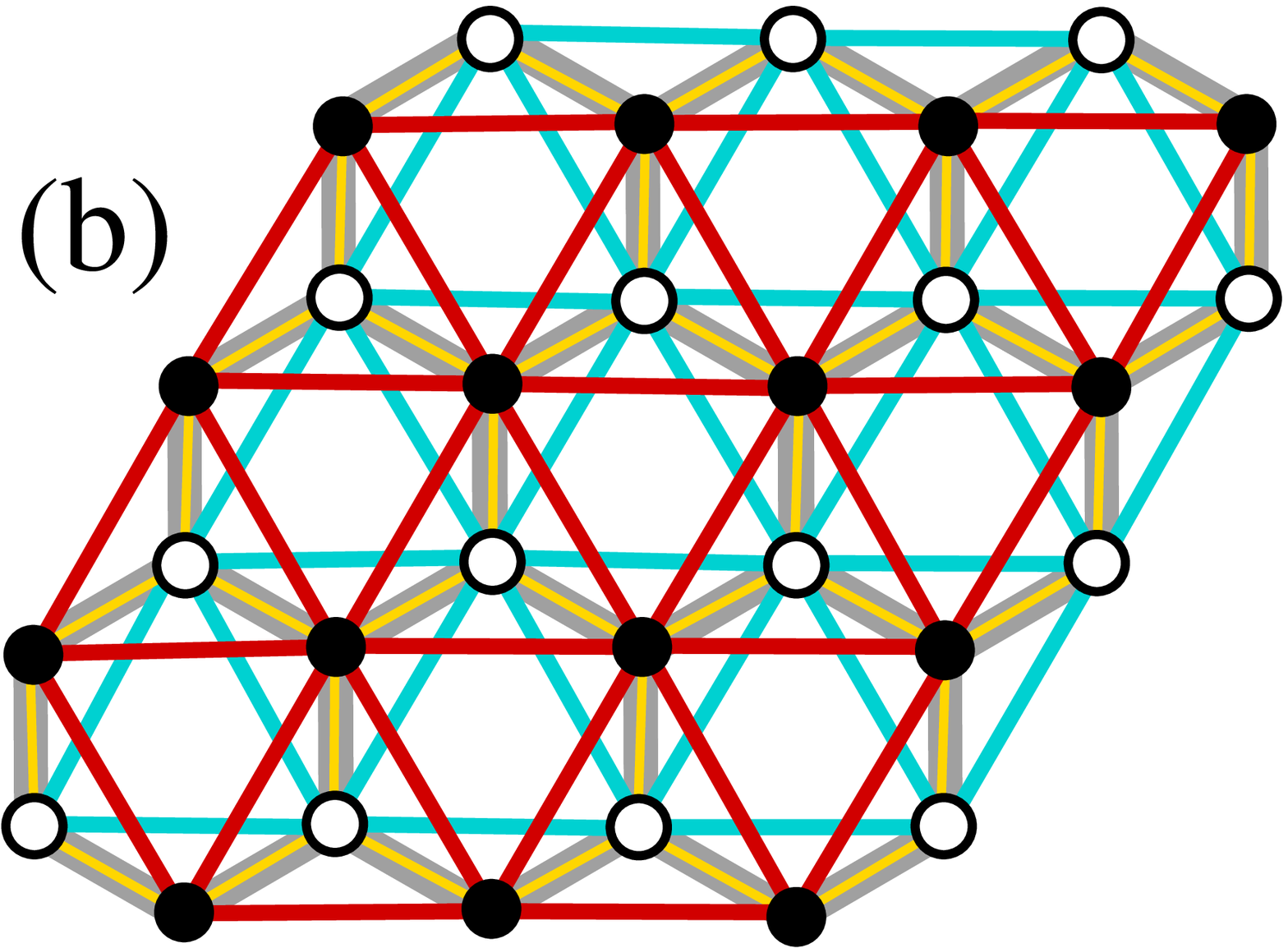}

\caption{\label{fig:Fig17} Schematic representation of a coarse-grained super-structure for a 
graphene sheet}
\end{figure}

\begin{figure}
\includegraphics[width=.5\textwidth]{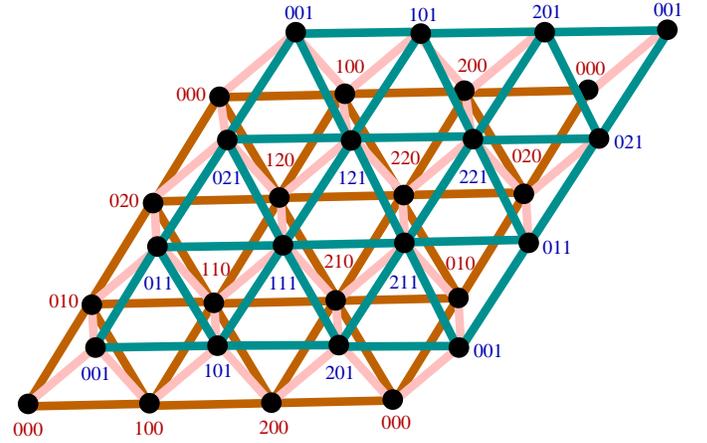}

\caption{\label{fig:Fig18} Illustration of the periodic dual-layer triangular lattice and labeling scheme;
blue indices refer to the upper layer, while red indices pertain to the lower plane.}
\end{figure}

Fig.~\ref{fig:Fig17} shows how the dual-layer systems is constructed as a caricature of the graphene 
lattice.  The image labeled (a) is a schematic illustration of a single hexagonal cell in a graphene
monolayer.  The bonding shares similarities with that in a benzene ring with delocalized $\pi$ orbitals
forming honeycomb networks of charge density above and below the plane occupied by the carbon atomic nuclei.  
The superimposed lattice work is a rigid network compatible with the symmetries of the graphene layer
and set up to capture the rigidity of the hexagonal cells making up a sheet of graphene.
With the honeycomb graphene pattern removed, the remaining lattice geometry and the labeling scheme for 
the crystal members is shown in Fig.~\ref{fig:Fig18}. 

With the superscript I representing the lower plane and II indicating the upper plane, the lattice
energy in the harmonic approximation is
\begin{align}
E \! = \! \frac{K}{2} \! \sum_{i,j=0}^{n-1} \! \left( \! \!  
\begin{array}{l}
\! \displaystyle \sum_{\alpha = \mathrm{I}}^{\mathrm{II}} \! 
\left \{ \! \! \! \begin{array}{l} \left [ \hat{x} \! \cdot \!  (\vec{\delta}_{i+1j}^{\alpha} 
- \vec{\delta}_{ij}^{\alpha} ) \! \right]^{2} \! \! + \! \left[ \hat{y} 
\! \cdot \! (\vec{\delta}_{ij+1}^{\alpha} -\vec{\delta}_{ij}^{\alpha}) \! \right]^{2} 
\\ + \left[ \frac{1}{\sqrt{2}}(\hat{x} + \hat{y}) \! \cdot \! (\vec{\delta}_{i+1j+1}^{\alpha} 
- \vec{\delta}_{ij}^{\alpha}) \!  \right]^{2} + \\ 
\left[ \frac{1}{\sqrt{2}} ( \hat{x} - \hat{y} ) \cdot (\vec{\delta}_{i+1j-1}^{\alpha} \! - \! \vec{\delta}_{ij}^{\alpha} ) \! \right ]^{2}
\end{array} \! \! \right   \}  \\
+ \kappa_{z} \left \{ \! \! \begin{array}{l} \left[ \hat{z} \cdot \left ( \vec{\delta}_{ij}^{\mathrm{II}} - 
\vec{\delta}_{ij}^{\mathrm{I}} \right ) \right ]^{2} \\ \! + \! 
\left[ \tfrac{1}{\sqrt{2}}(\hat{x} \! + \! \hat{z} ) \! \cdot \! 
\left( \vec{\delta}_{i+1j}^{\mathrm{II}} \! -\! \vec{\delta}_{ij}^{\mathrm{I}} \right) \right]^{2} \\ 
\! + \! \left[ \tfrac{1}{\sqrt{2}}(-\hat{x} \! + \! \hat{z} ) \! \cdot \!
\left( \vec{\delta}_{i-1j}^{\mathrm{II}} \! -\! \vec{\delta}_{ij}^{\mathrm{I}} \right) \! \right]^{2} \\ 
\! + \! \left[ \tfrac{1}{\sqrt{2}}(\hat{y} \! + \! \hat{z} ) \! \cdot \!
\left( \vec{\delta}_{ij+1}^{\mathrm{II}} \! -\! \vec{\delta}_{ij}^{\mathrm{I}} \right) \! \right]^{2} \\ 
\! + \! \left[ \tfrac{1}{\sqrt{2}}(-\hat{y} \! + \! \hat{z} ) \! \cdot \!
\left( \vec{\delta}_{ij-1}^{\mathrm{II}} \! -\! \vec{\delta}_{ij}^{\mathrm{I}} \right) \! \right]^{2} \end{array}  
\! \! \right \}
\end{array}
\! \!
\! \! \right)
\end{align}

The corresponding complex Hermitian 6$\times$6 matrix to be diagonalized has the form 
\begin{align}
\left[ \! \begin{array}{cccccc}
a_{\mathrm{I} x \mathrm{I} x} & a_{\mathrm{I} x \mathrm{I} y} &
a_{\mathrm{I} x \mathrm{I} z} & a_{\mathrm{I} x \mathrm{II} x} &
a_{\mathrm{I} x \mathrm{II} y} & a_{\mathrm{I} x \mathrm{II} z} \\
a_{\mathrm{I} y \mathrm{I}x} & a_{\mathrm{I} y \mathrm{I} y} &
a_{\mathrm{I} y \mathrm{I} z} & a_{\mathrm{I} y \mathrm{II} x} &
a_{\mathrm{I} y \mathrm{II} y} & a_{\mathrm{I} y \mathrm{II} z} \\
a_{\mathrm{I} z \mathrm{I} x} & a_{\mathrm{I} z \mathrm{I} y} &
a_{\mathrm{I} z \mathrm{I} z} & a_{\mathrm{I} z \mathrm{II} x} &
a_{\mathrm{I} z \mathrm{II} y} & a_{\mathrm{I} z \mathrm{II} z} \\
a_{\mathrm{II} x \mathrm{I}x} & a_{\mathrm{II} x \mathrm{I} y} &
a_{\mathrm{II} x \mathrm{I} z} & a_{\mathrm{II} x \mathrm{II} x} &
a_{\mathrm{II} x \mathrm{II} y} & a_{\mathrm{II} x \mathrm{II} z} \\
a_{\mathrm{II} y \mathrm{I} x} & a_{\mathrm{II} y \mathrm{I} y} &
a_{\mathrm{II} y \mathrm{I} z} & a_{\mathrm{II} y \mathrm{II} x} &
a_{\mathrm{II} y \mathrm{II} y} & a_{\mathrm{II} y \mathrm{II} z} \\
a_{\mathrm{II} z \mathrm{I}x} & a_{\mathrm{II} z \mathrm{I} y} &
a_{\mathrm{II} z \mathrm{I} z} & a_{\mathrm{II} z \mathrm{II} x} &
a_{\mathrm{II} z \mathrm{II} y} & a_{\mathrm{II} z \mathrm{II} z}
\end{array} \! \right ] = \left[ \! \begin{array}{cc} \hat{A} & \hat{B} \\ 
\hat{B}^{\dagger} & \hat{A} \end{array} \! \right]
\end{align}
where $\hat{A}$ and $\hat{B}$ are $3 \times 3$ matrices and $\hat{B}^{\dagger}$ is the 
Hermitian conjugate of $\hat{B}$.  The sub-matrices $\hat{A}$ and $\hat{B}$ are given by
\begin{align}
\hat{A} = \! K \left[ \! \! \!  \! \begin{array}{ccc}   \left( 
\! \! \begin{array}{c} \kappa_{z} \! + \! 4 \! - \! 2 \cos k_{x} \\  - 2 \cos k_{x} \cos k_{y} 
\end{array} \! \! \! \right)  \! \! \! \! & \! \! \! \! 2 \sin k_{x} \sin k_{y} \! \! \! \!
& \! \! \! \! 0
\\ \\ 2 \sin k_{x} \sin k_{y} \! \! \! \! & \! \! \! \! \left(
\! \! \begin{array}{c} \kappa_{z}\!  + \! 4 \!  - \! 2 \cos k_{y} \\  - 2 \cos k_{x} \cos k_{y}
\end{array} \! \! \! \right) \! \!\! \! & \! \! \! \!  0  \\ \\ 
0 \! \! \! \! & \! \! \! \! 0 \! \! \! \! & 3 \kappa_{z}  \end{array} \! \! \! \right]
\end{align}
where $\kappa_{z} \equiv K_{z}/K$
for $\hat{A}$ and 
\begin{align}
\hat{B} = \! K_{z} \left[ \! \! \! \begin{array}{ccc} -\cos k_{x}  \! & \!  0  \!  & \! 
\! -i\sin k_{x} \\ \\ 
0 \!  &  \! - \cos k_{y} \!  & \! -i \sin k_{y} \\ \\ -i \sin k_{x} \! & \! 
-i \sin k_{y} \!  &  \! - (1 \! + \! \cos k_{x} \! + \! \cos k_{y} ) \end{array} \! \! \right]
\end{align}
for $\hat{B}$
where the six eigenvalues for each wave-number pair are calculated numerically 
with code available in the EISPACK linear algebraic library for diagonalizing complex Hermitian 
matrices.   

On the other hand a locally stiff dual-layer system which may be regarded as a section of 
the three dimensional tetrahedral lattice such that the upper and 
lower layers are triangular lattices with connections between the layers.  
The dual-layer lattice structure based on the tetrahedral geometry is illustrated in 
Figure~\ref{fig:Fig18}.   The vertices of the upper layer are positioned above the centers of the 
triangles in the lower layer with bonds extending from atoms in the upper layer to each of 
the corners of the triangle below such that each atom in the dual-layer system is a 
member of a rigid tetrahedron; the result is a locally stiff layer, as in the dual-layer 
square lattice extended model, but with a very different geometric structure.

The lattice energy for the dual-layer tetrahedral system to quadratic order in the
displacements $\vec{\delta_{ij}}^{\mathrm{I}}$ and $\vec{\delta_{ij}}^{\mathrm{II}}$
has the form
\begin{align}
E =  \frac{K}{2} \! \! \sum_{i,j=0}^{n-1} \! \left( \! \! \!
\begin{array}{l}  \\  \! \displaystyle {\sum_{\alpha = \mathrm{I}}^{\mathrm{II}}}
\left \{ \! \! \!
\begin{array}{c} [ \hat{x} \! \cdot \!  (\vec{\delta}_{i+1j}^{\alpha} \! - \! \vec{\delta}_{ij}^{\alpha} ) \! ]^{2} + \\
\left [ \left( \tfrac{\hat{x}}{2} \! + \!
\tfrac{ \sqrt{3} \hat{y}}{2} \right) \! \cdot \! ( \vec{\delta}_{ij+1}^{\alpha} -
\vec{\delta}_{ij}^{\alpha} ) \right ]^{2} + \\
\left[ \left( \tfrac{\hat{x}}{2}  - \tfrac{ \sqrt{3} \hat{y}}{2} \right) \cdot ( \vec{\delta}_{i+1j-1}^{\alpha}
\! - \! \vec{\delta}_{ij}^{\alpha} ) \right]^{2} \end{array} \! \! \right \}
 \!+ \! \\ \\ \! \kappa_{z} \!
\left \{ \! \! \! \begin{array}{c}  \left[ \left( \tfrac{\hat{x}}{2} \!  + \! \tfrac{\hat{y}}{2 \sqrt{3}}
\! + \!
\sqrt{\tfrac{2}{3}} \hat{z}  \right) \cdot ( \vec{\delta}_{ij}^{\mathrm{II}}
- \vec{\delta}_{ij}^{\mathrm{I}} ) \right]^{2} + \\
\left[ \! \left(\!  - \tfrac{\hat{x}}{2} \!  + \! \tfrac{\hat{y}}{2 \sqrt{3}} \! + \!
\sqrt{\tfrac{2}{3}} \hat{z}
\right) \! \cdot \! ( \vec{\delta}_{ij}^{\mathrm{II}} \! - \! \vec{\delta}_{i+1j}^{\mathrm{I}})
 \! \right]^{2}  \! + \! \\
\left[ \! \left( \! -\tfrac{\hat{y}}{\sqrt{3}} \! + \! \sqrt{\tfrac{2}{3}} \hat{z}   \right)
\! \cdot \! ( \vec{\delta}_{ij}^{\mathrm{II}} \! - \! \vec{\delta}_{ij+1}^{\mathrm{I}} )
\! \right]^{2} \end{array}
\! \! \! \right \}  \end{array} \! \! \! \! \right)
\end{align}

Expressing the lattice energy in terms of Fourier components leads to
a $6 \times 6$ matrix to be diagonalized, which may again be written in terms of
$3 \times 3$ submatrices as $\left [ \begin{array}{cc} \hat{A} & \hat{B} \\
\hat{B}^{\dagger} & \hat{A} \end{array} \right]$, where
\begin{align}
\hat{A} = K \! \left[ \! \! \! \begin{array}{ccc} \left( \! \!
\begin{array}{c} \tfrac{\kappa_{z}}{2} \! + \! 3
\!- \! 2 \cos k_{x} \\ -\tfrac{1}{2} \cos (k_{y} \! - \! k_{x} )
\\ - \tfrac{1}{2} \cos k_{y}
\end{array} \! \! \! \right) \! \! & \! \! \tfrac{\sqrt{3}}{2} \left( \! \! \begin{array}{c}
\cos (k_{y} \! - \! k_{x}) \\ - \cos k_{y} \end{array} \! \! \right) \! \! & \! \! 0 \\ \\
\tfrac{\sqrt{3}}{2} \left( \! \! \begin{array}{c}
\cos (k_{y} \! - \! k_{x}) \\ - \cos k_{y} \end{array} \! \! \right) \! \!
& \! \! \left( \! \!
\begin{array}{c} \tfrac{\kappa_{z}}{2} \! + \! 3 \! - \! \tfrac{3}{2} \cos (k_{y} ) \\
  - \tfrac{3}{2} \cos (k_{y} -k_{x} )  \end{array} \! \! \!\right) \! \! & \! \! 0  \\ \\
0 \! \! & \! \! 0 \! \! &  \! \! 2  \kappa_{z} \end{array}  \! \! \right]
\end{align}
where again $\kappa_{z} \equiv K_{z}/K$
for the sub-matrix $\hat{A}$ and
\begin{align}
\hat{B} \! = \! \tfrac{K_{z}}{4} \! \! \left[ \! \! \! \begin{array}{ccc} -(1 \!+ \! e^{-i k_{x}} \!)    &
\tfrac{1}{\sqrt{3}} \! (e^{-ik_{x}}\! - \! 1 \! ) \! \!  &  \! \! \sqrt{\tfrac{8}{3}} (e^{-ik_{x}} \! - \! 1 \!
) \\ \\
\tfrac{1}{\sqrt{3}} (e^{-ik_{x}} \! - \! 1 \!)  & -\tfrac{1}{3} \! \! \left ( \! \!
\begin{array}{c} 1 \! + \! e^{-ik_{x}}  \\ + 3 e^{-ik_{y}} \end{array} \! \! \! \right )
\! \!  & \!  \! \tfrac{\sqrt{8}}{3} \! \! \left ( \! \! \begin{array}{c} 2 e^{-ik_{y}} - \\ e^{-i k_{x}} -1
\end{array} \! \! \! \right ) \\ \\ \sqrt{\tfrac{8}{3}} (e^{-i k_{x}} \! - \! 1 \!)  &  \frac{\sqrt{8}}{3}
\! \left( \! \! \begin{array}{c} 2 e^{-i k_{y}} - \\ e^{-i k_{x}} \! - \! 1 \end{array} \! \! \! \right)
\! \! & \! \! -\tfrac{8}{3} \! \! \left( \! \! \begin{array}{c} 1 \! + \!
e^{-ik_{z}} \\ - \! e^{-ik_{y}} \end{array} \! \! \! \right )
\end{array} \! \! \! \! \right]
\end{align}
for the sub-matrix $\hat{B}$.

The use of a dual-layer lattice geometry to provide resistance to transverse deviations   
is not sufficient to prevent a rapid divergence in $\delta_{\mathrm{RMS}}$ with 
increasing system size $L$.  Whereas thermally averaged mean square fluctuations grew very
slowly (i.e. logarithmically) when the atomic motions are confined to the lattice plane,
$\delta_{\mathrm{RMS}}$ for the dual-layer systems increases linearly with $L$; ultimately, it
is not difficult for the sheet to bend and flex in the presence of thermal fluctuations 
in spite of its locally stiff characteristics.  The diverging mean square deviations from 
equilibrium and other thermodynamic characteristics of the the dual-layer square lattice in the 
extended coupling scheme and its counterpart based on a tetrahedral geometry are   
examined, with consideration given to the effects of increasing $L$ and variations in 
the inter-layer coupling strength.

\begin{figure}
\includegraphics[width=.49\textwidth]{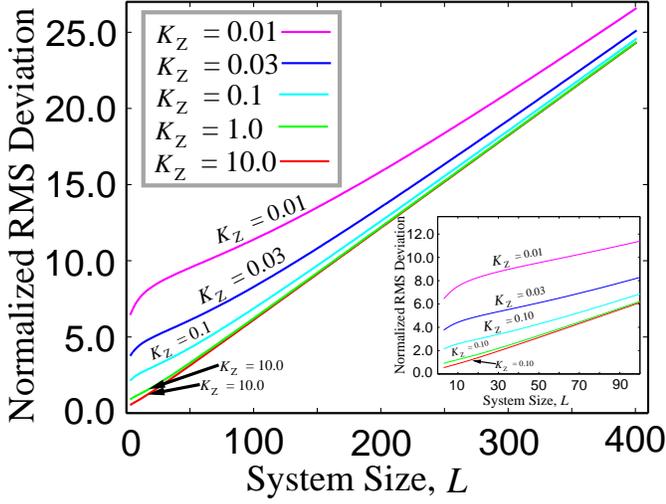}

\caption{\label{fig:Fig20} Normalized mean square displacements
for $K_{z}$ = 0.01, 0.03, 0.1, 1.0, and 10.0 for the dual-layer 
square lattice with an extended coupling scheme, where 
$K_{z}$ is in units of the intralayer coupling $K$.  
The inset is a closer view
of the RMS curves.}
\end{figure}

The graph of the normalized RMS displacement, shown in Fig.~\ref{fig:Fig20}
 for the dual-layer square lattice with an 
extended coupling pattern shows a dependence on systems size which is an asymptotically 
linear growth in $L$.
$\delta_{\mathrm{RMS}}$ 
curves for several values of the interlayer coupling 
$K_{z}$ are shown; the intra-layer coupling is taken to be unity, so 
$K_{z}$ is effectively expressed in units of $K$.  Although the relative interlayer couplings range over three orders       
magnitude, there is little variation of the curves, especially for $K_{z} = 
\left \{ 0.1, 1.0, 10.0 \right \}$.
Similarly, the $\delta_{\mathrm{RMS}}$ curves ultimately vary linearly in the system size 
$L$ with little dependence on the relative magnitude of $K_{z}$, which again is expressed in 
units of $K$.

Again, it is useful to examine the density of states for the eigenvalues in the case of 
the locally rigid dual-layer systems, which are richer than the density of states profiles 
corresponding to rigid three dimensional lattices or those of the single layer geometries with
atomic fluctuations confined to intra-planar motion. Although details of the density of states profiles for 
the two geometries differ, the both curves show a divergence of the density of states with 
decreasing eigenvalue magnitude, whereas the density states remained constant in the case of 
the planar systems with exclusively intra-planar motion and vanished altogether for 
the rigid three dimensional systems.  Inset (a) of Fig.~\ref{fig:Fig19} show the DOS 
for the dual-layer square lattice, while inset (b) is a graph of the density of states for the 
locally stiff tetrahedral system.  The DOS cusp for both lattices at the zero eigenvalue point 
is responsible for the rapid divergence of $\delta_{\mathrm{RMS}}$ with increasing $L$.

\begin{figure}
\includegraphics[width=.49\textwidth]{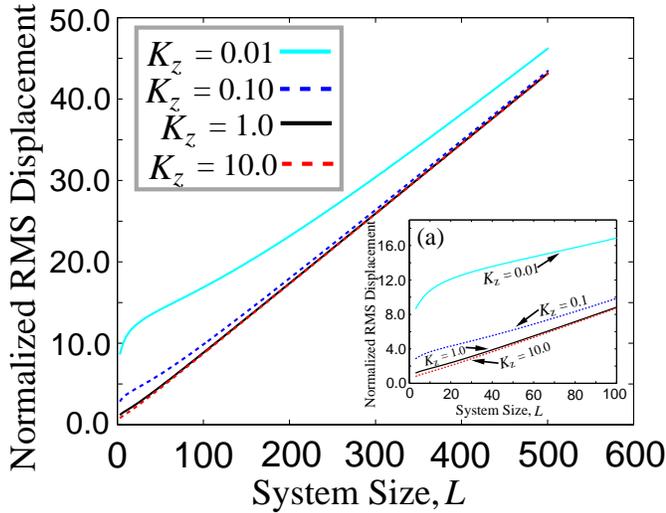}

\caption{\label{fig:Fig21} Normalized mean square displacements 
for $K_{z}$ = 0.01, 0.1, 1.0, and 10.0 for the dual-layer tetrahedral
lattice where $K_{z}$ is in units of the triangular lattice intra-layer 
coupling $K$.  The inset is a closer view 
of the RMS curves.} 
\end{figure}

\begin{figure}
\includegraphics[width=.49\textwidth]{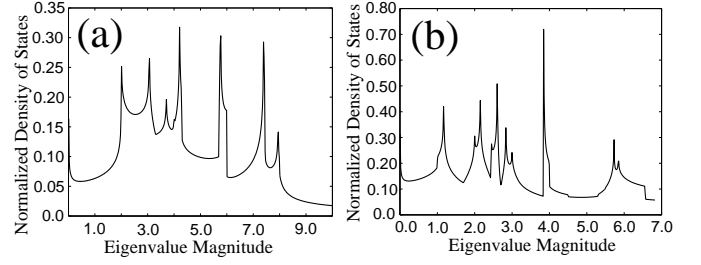}

\caption{\label{fig:Fig19} Normalized Eigenvalue Density of States for the
dual-layer cubic system with with nearest and next-nearest
neighbor interactions (a) and for the dual-layer locally 
stiff lattice based on a tetrahedral lattice geometry (b) 
for system size $L = 5001$.}
\end{figure}

\begin{figure}
\includegraphics[width=.49\textwidth]{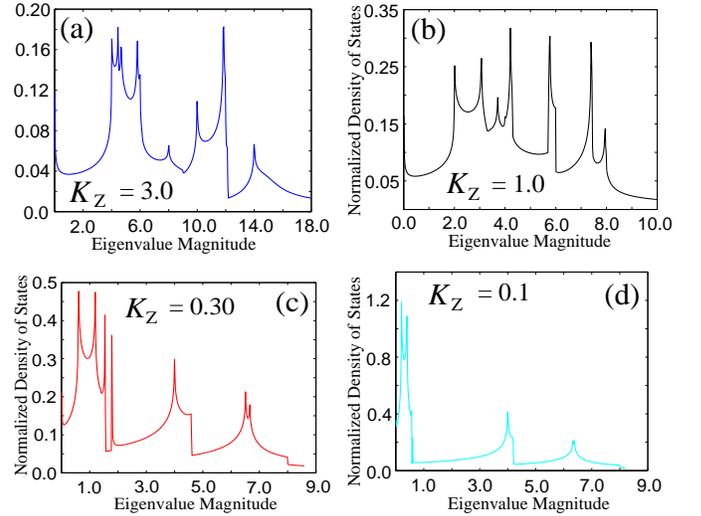}

\caption{\label{fig:Fig22} Normalized density of states profiles
for $K_{z}$ = 3.0, 1.0, 0.3, and 0.1 for the dual-layer square
lattice with an extended coupling scheme
where $K_{z}$ is in units of the triangular lattice intra-layer
coupling $K$.}
\end{figure}

\begin{figure}
\includegraphics[width=.49\textwidth]{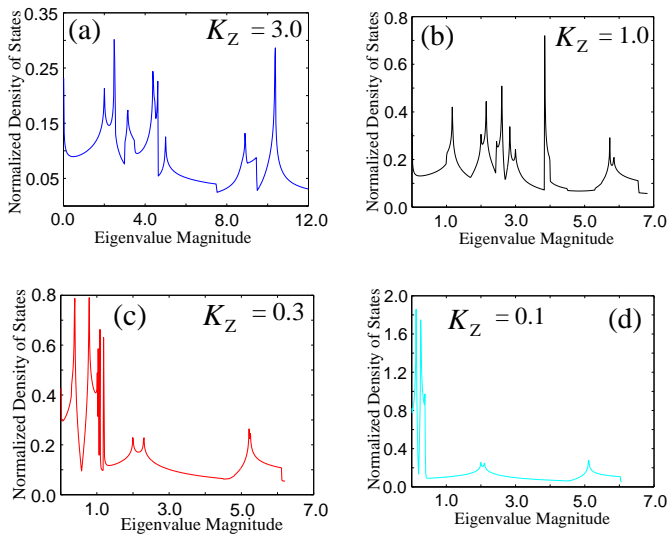}

\caption{\label{fig:Fig23} Normalized density of states profiles
for $K_{z}$ = 3.0, 1.0, 0.3, and 0.1 for the dual-layer tetrahedral
lattice where $K_{z}$ is in units of the triangular lattice intra-layer
coupling $K$.}
\end{figure}

While adjusting the interaction between the layers to enhance the 
resistance to local transverse perturbations has little effect on the 
mean square fluctuations for large values of $L$, the eigenvalue density of states   
evolves as the interplanar to intraplanar coupling ratio $\kappa_{z} = K_{z}/K$
is modified.  Density of states profiles for $\kappa_{z}$ values ranging from  $\kappa_{z} = 0.1$ to 
$\kappa_{z} = 3.0$ are shown for the dual-layer square system with and extended 
coupling pattern in Fig.~\ref{fig:Fig22} and for the tetrahedral counterpart in 
Fig.~\ref{fig:Fig23}.

Density of states profiles are shown for strong ($\kappa_{z} = 3.0$) and moderate
($\kappa_{z} = 1.0$) values of the the coupling ratio in insets (a) and (b) of Fig.~\ref{fig:Fig22} and 
Fig.~\ref{fig:Fig23}, and there is little change in the DOS curve in the low eigenvalue regime. 
On the other had, as $\kappa_{z}$ decreases further and the interplanar coupling begins to fall below     
parity with that in the plane, the eigenvalue density of states profiles begin to change 
more drastically, as may be seen in panels (b) and (c) of Fig.~\ref{fig:Fig22} for the dual-layer
square system and Fig.~\ref{fig:Fig23} for the dual-layer tetrahedrally based geometry; the 
distribution in both cases rapidly grows narrower with decreasing $\kappa_{z}$.
Although the two lattice geometries are very distinct, similar (and likely generic to
locally rigid dual-layer lattices) trends may be seen in the 
DOS profiles in the regime of low eigenvalues as $\kappa_{z}$ is reduced.

\section{Coupling to a Substrate}

  We incorporate an attractive interaction with a flat substrate  by including an additional harmonic potential acting on the lower members
of the tetrahedral and square extended coupling dual-layer systems.   We take the attraction to depend only on the atomic shift 
$\delta^{\mathrm{I}z}_{ij}$ above the planar system, and the additional term hence has the form $\frac{K_{s}}{2} \left( \delta_{ij}^{Iz}
\right)^{2}$ 

\begin{figure}
\includegraphics[width=.49\textwidth]{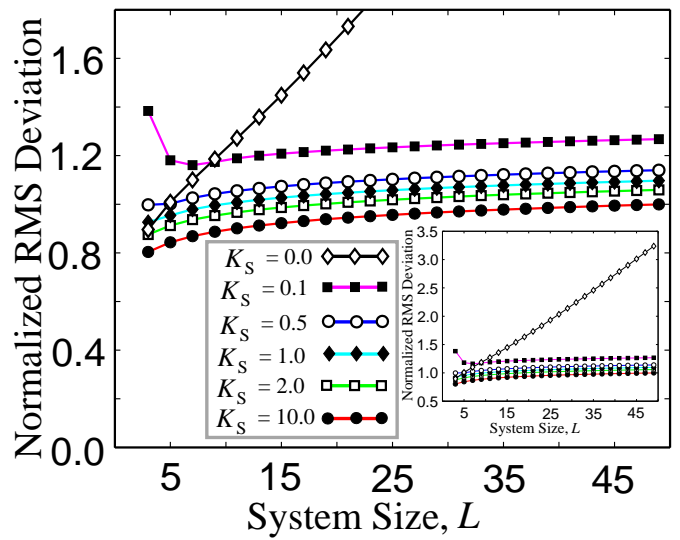}

\caption{\label{fig:Fig24} For the dual-layer square lattice in the 
extended scheme, the 
main figure and the inset are graphs of mean square
fluctuations versus system size $L$ for a range of substrate couplings $K_{s}$, with $K_{s}$
given in units of the inter-planar and intra-planar coupling constant                  
(both equal to $K$).  The symbol legend on the main plot also pertains to the inset}
\end{figure}

\begin{figure}
\includegraphics[width=.49\textwidth]{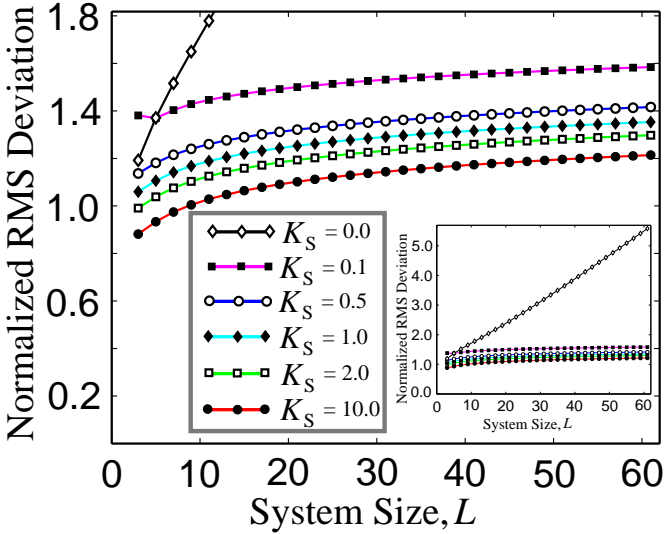}

\caption{\label{fig:Fig25} For the dual-layer tetrahedral lattice, the 
main figure and the inset are graphs of mean square 
fluctuations versus system size $L$ for a range of substrate couplings $K_{s}$, with $K_{s}$ 
given in units of the inter-planar and intra-planar coupling constant  
(both equal to $K$).  The symbol legend on the main plot also pertains to the inset}
\end{figure}

Figure~\ref{fig:Fig24} shows the effect of the substrate coupling on $\delta_{\mathrm{RMS}}$ in the case of 
the dual-layer square system with an extended coupling scheme. The 
graph, which shows mean square deviation curves for a wide range of $K_{s}$ values, indicates the capacity of even a 
very mild substrate coupling to suppress thermally induced undulations in the dual-layer sheet.  
Similarly, for the tetrahedrally based dual-layer lattice geometry, an attractive interaction with a substrate 
considerably reduces fluctuations transverse to the lattice planes, preventing a rapid divergence of 
$\delta_{\mathrm{RMS}}$.  The mean square deviation curves are shown in Figure~\ref{fig:Fig25}.

\begin{figure}
\includegraphics[width=.49\textwidth]{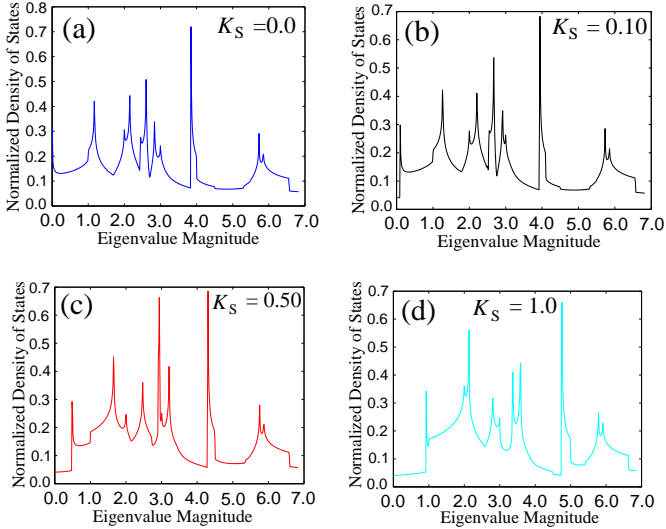}

\caption{\label{fig:Fig26} Normalized Density of States for the Dual layer system for various
substrate coupling strengths $K_{s}$.  Panels (a), (b), (c), and (d) show the density of states 
for $K_{s}$ equal to $0.0$, $0.1$, $0.5$, and $1.0$ respectively.}
\end{figure}

We also examine the effect of an attractive substrate coupling on the density of states profiles, and 
results are displayed in Fig.~\ref{fig:Fig26} for a range of substrate coupling constants $K_{s}$.  
With increasing $K_{s}$, a salient trend is the opening of a separation between the sharp cusp and the 
zero eigenvalue mark on the abscissa.  The migration of the maximum formerly at the zero eigenvalue point to 
a peak at a larger eigenvalue is associated with a sharp reduction of the mean square fluctuations about 
equilibrium, and the lattice is better able to withstand transverse fluctuations.   

The presence of a flat substrate plays a very important role in dictating the overall structure and
amplitude of ripples in the dual-layer geometries we report on here.  This result is in accord
with recent experiments on graphene sheets deposited on cleaved mica substrates~\cite{Lui}, 
where the careful preparation of flat substrates significantly 
dampens the ripple amplitude, whereas much larger undulations are seen with sheets attached to substrates 
with poorer control over the flatness.

\begin{figure}
\includegraphics[width=.48\textwidth]{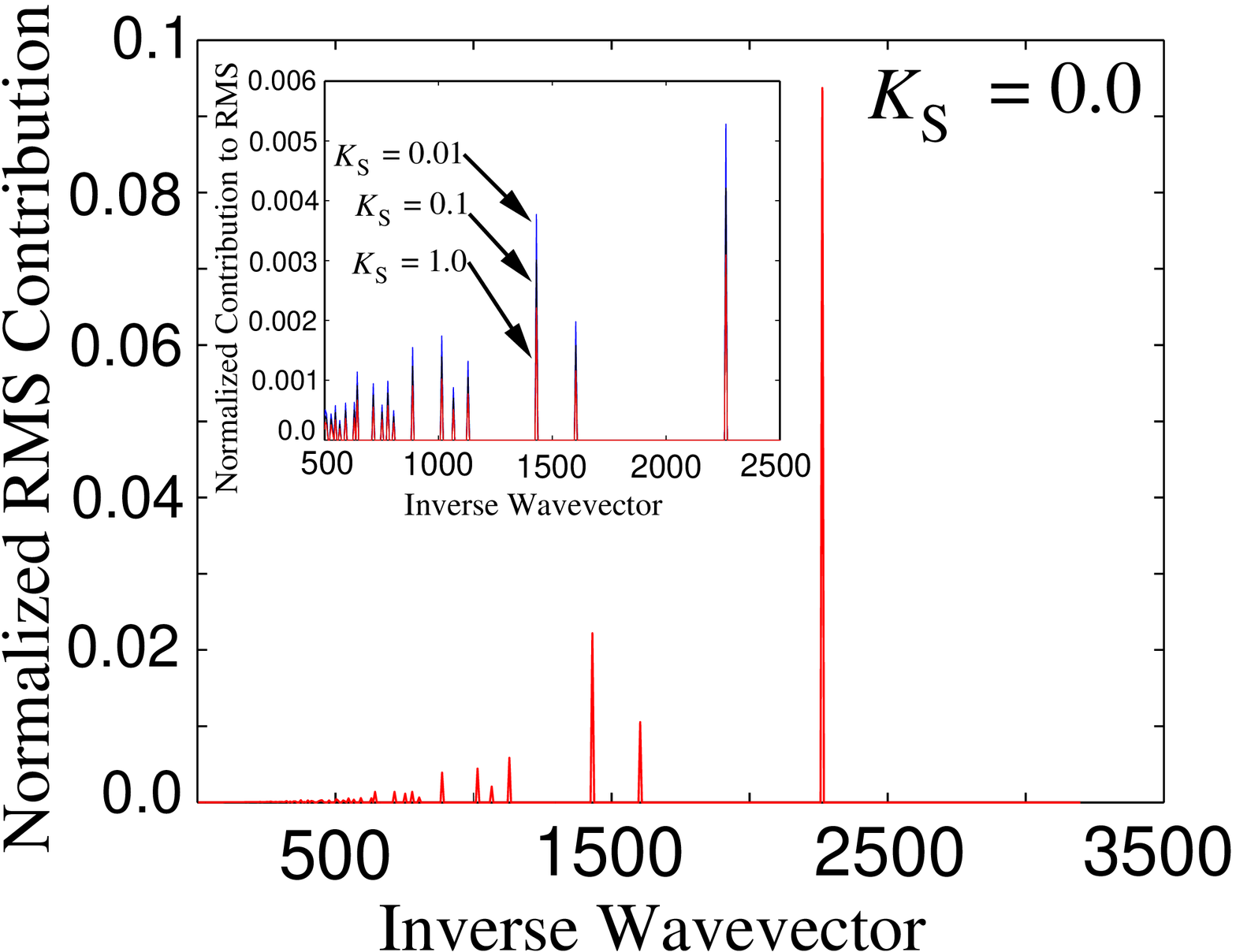}

\caption{\label{fig:Fig27} Normalized Eigenvalue Density of States for the
dual-layer triangular system with system size $L = 5001$.}
\end{figure}

\begin{figure}
\includegraphics[width=.48\textwidth]{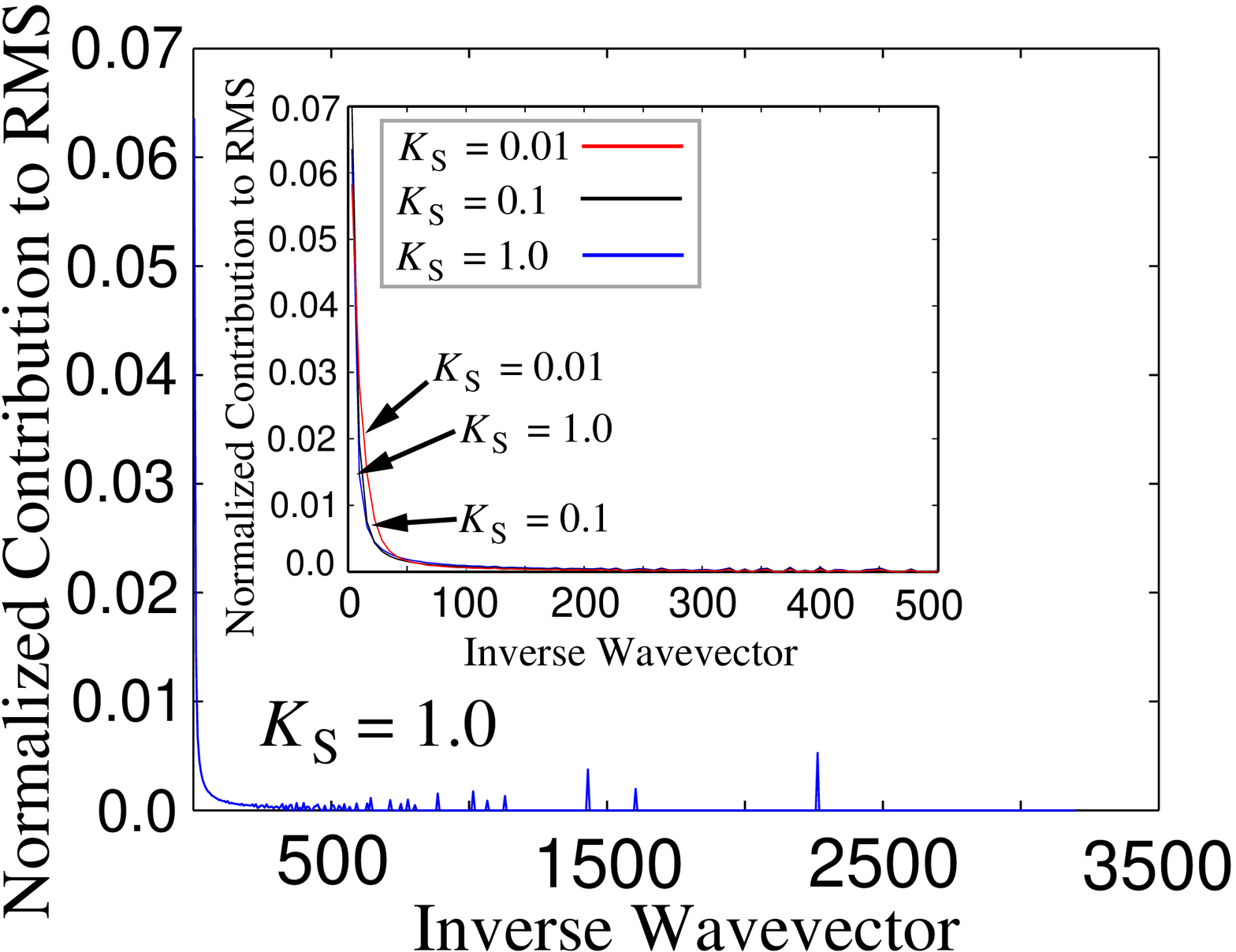}

\caption{\label{fig:Fig28} Normalized Eigenvalue Density of States for the
dual-layer triangular system with system size $L = 5001$.}
\end{figure}

To determine which length scales are associated with the strongest contributions to the thermally 
averaged mean square 
deviations about equilibrium, we have prepared histograms showing the relative contribution to 
$\delta_{\mathrm{RMS}}$ versus inverse wave-vector magnitude, with the latter providing a length scale.
Apart from a significant diminution in the height of thermally excited undulations in the 
dual-layered sheets, we also find a considerable reduction in their typical wavelength.
In figure~\ref{fig:Fig27}, for the dual-layer tetrahedral lattice in the absence of a substrate couplings, the dominant    
contribution to $\delta_{\mathrm{RMS}}$ comes from large length scales comparable to the scale of the 
lattice.  However the picture changes with the activation of a finite substrate coupling as may be seen in the 
inset with the peak height at the minimal wave-number decreasing with increasing $K_{s}$.  
Moreover, as may be seen in Figure~\ref{fig:Fig28}, introducing even a weak anchoring to the foundation below immediately creates a           
strong peak in the short wavelength regime, skewing the size of thermally induced ripples toward smaller length scales.

    In conclusion, we have examined thermally induced fluctuations about 
equilibrium in two and three dimensional crystalline solids with a local 
bonding scheme.  While long-range crystalline order may exist in three dimensional
crystal lattices, some geometries (e.g. the simple cubic lattice) are not    
rigid when only nearest-neighbor couplings are taken into account, and an  
extended coupling scheme is needed to prevent the divergence of mean square 
fluctuations with increasing system size $L$. 

  In two dimensional lattices, we find RMS fluctuations to increase at a very slow 
(logarithmic) rate when motion is confined to the lattice plane.  On the other hand,
when transverse motion is permitted, thermal fluctuations are very effective in 
bringing about significant vertical displacements of particles which contribute to rapidly growing 
deviations from equilibrium, and $\delta_{\mathrm{RMS}}$ ultimately diverges at a linear 
rate in $L$.  The asymptotically linear divergence in the mean square deviations from 
equilibrium is insensitive to the strength of the interlayer coupling; $\delta_{\mathrm{RMS}}$ 
values appear to converge and eventually show identical behavior 
with increasing system size whether the coupling $K_{z}$ established  
between the layers to provide local stiffness is quite weak or very strong relative to the 
bonding $K$ between atoms in the same layer.

  Introducing a coupling $K_{s}$ to a flat substrate very effectively hinders transverse fluctuations in 
two dimensional crystal lattices, even in the coupling is very weak, and reflects the 
importance of a substrate in shaping the characteristics of ripples set up by thermal 
fluctuations by inhibiting transverse deviations.  An 
attractive coupling to a fixed substrate also reduces the typical lateral 
length scale or wavelength of thermally excited undulations in lattices bound to a 
substrate.  These tendencies are consistent with recent experimental observations 
that  control
over the flatness of the underlying surface is directly related to the amplitude and length 
scale of thermally induced ripples. 
 
\begin{acknowledgments}

\end{acknowledgments}


\end{document}